\documentclass[sigconf,table]{acmart}

\usepackage{amsmath,mathtools}
\usepackage{colortbl}
\usepackage{algorithm2e}
\usepackage{multirow}
\usepackage{pgf}
\usepackage{tikz}
\usepackage{xcolor}
\usepackage{nicematrix}
\usetikzlibrary{arrows,automata,fit,calc, positioning}
\usepackage{soul}
\usepackage[makeroom]{cancel}

\setcopyright{acmcopyright}
\copyrightyear{2018}
\acmYear{2018}
\acmDOI{10.1145/1122445.1122456}

\acmConference[Woodstock '18]{Woodstock '18: ACM Symposium on Neural
  Gaze Detection}{June 03--05, 2018}{Woodstock, NY}
\acmBooktitle{Woodstock '18: ACM Symposium on Neural Gaze Detection,
  June 03--05, 2018, Woodstock, NY}
\acmPrice{15.00}
\acmISBN{978-1-4503-XXXX-X/18/06}

\begin{document}

\title{GLL-based Context-Free Path Querying for Neo4j}

\author{Vadim Abzalov}
\affiliation{
		\institution{Saint Petersburg State University}
		\city{St. Petersburg}
		\country{Russia}
	}
\email{vadim.i.abzalov@gmail.com}

\author{Vlada Pogozhelskaya}
\affiliation{%
  \institution{National Research University \\ Higher School of Economics}
  \streetaddress{16 Soyuza Pechatnikov Street}
  \city{St. Petersburg}
  \country{Russia}
  \postcode{190121}
}
\email{pogozhelskaya@gmail.com}

\author{Vladimir Kutuev}
\affiliation{
		\institution{Saint Petersburg State University}
		\city{St. Petersburg}
		\country{Russia}
	}
\email{vladimir.kutuev@gmail.com}

\author{Semyon Grigorev}
\affiliation{
		\institution{Saint Petersburg State University}
		\streetaddress{7/9 Universitetskaya nab.}
		\city{St. Petersburg}
		\country{Russia}
		\postcode{199034}
	}
\email{s.v.grigoriev@spbu.ru}
\orcid{0000-0002-7966-0698}

\renewcommand{\shortauthors}{Abzalov, Pogozhelskaya, Kutuev, Grigorev.}
\newcommand{\mycomment}[1]{}

\begin{abstract}
  We propose GLL-based \textit{context-free path querying} algorithm which handles queries in Extended Backus-Naur Form (EBNF) using Recursive State Machines (RSM). Utilization of EBNF allows one to combine traditional regular expressions and mutually recursive patterns in constraints natively. The proposed algorithm solves both the \textbf{reachability-only} and the \textbf{all-paths} problems for the \textbf{all-pairs} and the \textbf{multiple sources} cases. The evaluation on real-world graphs demonstrates that utilization of RSMs increases performance of query evaluation. Being implemented as a stored procedure for Neo4j, our solution demonstrates better performance than a similar solution for RedisGraph. Performance of our solution of regular path queries is comparable with performance of native Neo4j solution, and in some cases our solution requires significantly less memory.
\end{abstract}

    \ccsdesc[500]{Information systems~Graph-based database models}
	\ccsdesc[500]{Information systems~Query languages for non-relational engines}
	\ccsdesc[500]{Theory of computation~Grammars and context-free languages}
    \ccsdesc[300]{Mathematics of computing~Paths and connectivity problems}

\keywords{Graph database, context-free path querying, CFPQ, reachability problem, all-paths problem, generalized LL, GLL}
\maketitle

\section{Introduction}

Context-free path querying (CFPQ) allows one to use context-free grammars to specify constraints on paths in edge-labeled graphs. Context-free constraints are more expressive than the regular ones (RPQ) and can be used for graph analysis in such areas as bioinformatics (hierarchy analysis~\cite{10.1007/978-3-319-46523-4_38}, similarity queries~\cite{SubgraphQueriesbyContextfreeGrammars}), data provenance analysis~\cite{8731467}, static code analysis~\cite{Zheng, 10.1145/373243.360208}. Although a lot of algorithms for CFPQ were proposed, poor performance on real-world graphs and bad integration with real-world graph databases and graph analysis systems still are problems which hinder the adoption of CFPQ.

The problem with the performance of CFPQ algorithms in real-world scenarios was pointed out by Jochem Kuijpers~\cite{Kuijpers:2019:ESC:3335783.3335791} as a result of an attempt to extend Neo4j graph database with CFPQ. Several algorithms, based on such techniques as LR parsing algorithm~\cite{10.1007/978-3-319-91662-0_17}, dynamic programming~\cite{hellingsRelational}, LL parsing algorithm~\cite{10.1145/3167132.3167265}, linear algebra~\cite{Azimov:2018:CPQ:3210259.3210264}, were implemented using Neo4j as a graph storage and evaluated. None of them was performant enough to be used in real-world applications.   

Since Jochem Kuijpers pointed out the performance problem, it was shown that linear algebra based CFPQ algorithms, which operate over the adjacency matrix of the input graph and utilize parallel algorithms for linear algebraic operations, demonstrate good performance~\cite{10.1145/3398682.3399163}. Moreover, the matrix-based CFPQ algorithm is a base for the first full-stack support of CFPQ by extending RedisGraph graph database~\cite{10.1145/3398682.3399163}.  

However, adjacency matrix is not the only possible format for graph representation, and data format conversion may take a lot of time, thus it is not applicable in some cases. As a result, the development of a performant CFPQ algorithm for graph representations not based on matrices and its integration with real-world graph databases is still an open problem. Moreover, while the \textbf{all pairs context-free constrained reachability} is widely discussed in the community, such practical cases as the \textbf{all-paths} queries and the \textbf{multiple sources} queries are not studied enough.

Additionally, almost the all existing solutions provide algorithms for \textbf{reachability-only} problem. Recently, Jelle Helligs in~\cite{10.1007/978-3-030-61133-0_7} and Rustam Azimov in~\cite{10.1145/3461837.3464513} proposed algorithms which allow one to extract paths satisfied specified context-free constraints. The ability to extract paths of interest is important for some applications where the user wants to know not only the fact that one vertex is reachable from another, but also to get a detailed explanation why this vertex is reachable. One of such application is a program code analysis where the fact is a potential problem in code, and paths can help to analyze and fix this problem. While utilization of general-purpose graph databases for code analysis gaining popularity~\cite{9064792}, CFPQ algorithms which provide structural representation of paths are not studied enough. 

It was shown that the generalized LL (GLL) parsing algorithm can be naturally generalized to CFPQ algorithm~\cite{Grigorev:2017:CPQ:3166094.3166104}. Moreover, this provides a natural solution not only for the \textbf{reachability-only} problem but also for the \textbf{all-paths} problem. At the same time, there exists a high-performance GLL parsing algorithm~\cite{10.1007/978-3-662-46663-6_5} and its implementation in Iguana project\footnote{Iguana parsing framework: \url{https://iguana-parser.github.io/}. Accessed: 12.11.2021.}. 

Additionally, pure context-free grammars in Backus-Naur Form (BNF) are too verbose to express complex constraints. But almost the all algorithms require query to be in such form. At the same time, Extended Backus-Naur Form (EBNF) can be used to specify context-free languages. EBNF allows combining typical regular expressions with mutually recursive rules which required to specify context-free languages. That does EBNF more user-friendly. But there are no CFPQ algorithms that utilize EBNF for query specification.

In this paper, we generalize the GLL parsing algorithm to handle queries in EBNF without its transformation. We show that it allows to increase performance of query evaluation. Also integrate our solution with the Neo4j graph database and evaluate it. So, we make the following contributions in the paper.
\begin{itemize}
    \item We propose the GLL-based CFPQ algorithm which can handle queries in Extended Backus-Naur Form without transformation. Our solution utilizes Recursive State Machines (RSM)~\cite{10.1145/1075382.1075387} for it. The proposed algorithm can be used to solve both \textbf{reachability-only} and \textbf{all-paths} problems.
    \item We provide an implementation of the proposed CFPQ algorithm. By experimental study on real-world graphs we show that utilization of RSM-s allows one to increase performance of query evaluation. 
    \item We integrate the implemented algorithm with Neo4j by providing a respective stored procedure that can be called from Cypher. Currently, we use Neo4j as a graph storage and do not extend Cyper to express context-free path patterns. So, expressive power of our solution is limited: we can not use full power of Cypher inside our constrains. Implementing a query language extension amounts to a lot of additional effort and is a part of future work.
    \item We evaluate the proposed solution on several real-world graphs. Our evaluation shows that the proposed solution in order of magnitude faster than similar linear algebra based solution for RedisGraph. Moreover, we show that our solution is compatible with native Neo4j solution for RPQs, and in some cases requires significantly less memory. Note that while Cypher expressively is limited, our solution can handle arbitrary RPQs.
\end{itemize}

The paper has the following structure. In section~\ref{section:preliminaries} we introduce basic notions and definitions from graph theory and formal language theory. Then, in section~\ref{section:RSM} we introduce \textit{recursive state machines} and provide example of CFPQ evaluation using na{\"i}ve strategy which may leads to infinite computations. Section~\ref{section:GLL-CFPQ} contains description GLL-based CFPQ evaluation algorithm which uses RSM and solves problems of na{\"i}ve strategy from the previous section: the algorithm always terminates, can build finite representation of all paths of interest. Section~\ref{section:dataset} introduces a data set (both graphs and queries) which will be used further for experiments. Further, in section~\ref{section:BFN_or_RSM} we use this data set to compare different versions of GLL-based CFPQ algorithm. After that, in section~\ref{section:cfpq_fro_neo4j} we provide details on integration of the best version into Neo4j graph database, and evaluate our solution on the data set introduced before. Related work discussed in section~\ref{section:related_work}. Final remarks and conclusion provided in section~\ref{section:conclusion}.
\section{Preliminaries}\label{section:preliminaries}

In this section, we introduce basic definitions and notations for graphs, regular languages, context-free grammars, and finally, we formulate variants of the FLPQ problem.

We use a directed edge-labelled graph as a data model. We denote a graph as $D=\langle V,E,L \rangle$, where $V$ is a finite set of vertices, $E \subseteq V \times L \times V$ is a set of edges, and $L$ is a set of edge labels. A path $\pi$ in a graph $D$ is a sequence of edges: $u \xrightarrow{l_0} \ldots \xrightarrow{l_m} v$. We denote a path between vertices $u$ and $v$ as $u\pi v$. The function $\omega$ maps a path to a word by concatenating the labels of its edges:
\begin{equation}\label{eqn:omega}
  \omega(\pi) = \omega(u \xrightarrow{l_0} \ldots \xrightarrow{l_m} v) = l_0\ldots l_m .  
\end{equation}

In the context of formal languages, we would like to remind the reader of some fundamental facts about regular expressions and regular languages.

\begin{definition}
    A \textit{regular expression} over the alphabet $\Sigma$ specifies a \textit{regular language} using the following syntax.
    \begin{itemize}
        \item If $t \in \Sigma$, then $t$ is a regular expression.
        \item The concatenation $E_1 \cdot E_2$ of two regular expressions $E_1$ and $E_2$ is a regular expression. Note that in some cases $\cdot$ may be omitted.
        \item The union $E_1 \mid E_2$ of two regular expressions $E_1$ and $E_2$ is a regular expression.
        \item The Kleene star $E^* = \bigcup_{n=0}^{n=\infty} R^n$ of a regular expression $E$ is a regular expression.
    \end{itemize}
\end{definition}

In some cases $E^+$ can be used as syntactic sugar for $E \cdot E^*$. 

Both regular expressions and finite automata specify regular languages and can be converted into each other. Every regular language can be specified using deterministic finite automata without $\varepsilon$-transitions. Note that regular languages form a strong subset of context-free languages.

\begin{definition}\label{def:bnf}
A \textit{context-free grammar} is a 4-tuple $G=\langle \Sigma, \mathcal{N}, P, S\rangle$, where $\Sigma$ is a finite set of terminals, $\mathcal{N}$ is a finite set of nonterminals, $S \in N$ is the start nonterminal, and $P$ is a set of productions. Each production has the following form: $N_i \to w$, where $N_i \in \mathcal{N}$ is the left-hand side of the production, and $w \in (\Sigma \cup \mathcal{N} )^*$ is the right-hand side of the production.    
\end{definition}
 
 For simplicity, $S \to w_1 \mid w_2$ is used instead of $S \to w_1; S \to w_2$. Some example grammars are presented in equations~\ref{eqn:g_1},~\ref{eqn:g_2}, and~\ref{eqn:geo}. 

 \begin{definition}
 A \textit{derivation step} (in the grammar $G =\langle \Sigma, \mathcal{N}, P, S\rangle$) is a production application: having a sequence of form $w_1N_0w_2$, where $N_0 \in \mathcal{N}$ and $w_1, w_2 \in (\Sigma \cup \mathcal{N})^*$, and a production $N_0 \to w_3$, one gets a sequence $w_1w_3w_2$, by replacing the left-hand side of the production with its right-hand side.    
 \end{definition}

\begin{definition}
A word $w \in \Sigma^*$ is \textit{derivable} in the grammar $G=\langle \Sigma, \mathcal{N}, P, S\rangle$ if there is a sequence of derivation steps such that the initial sequence is the start nonterminal of the grammar and $w$ is a final sequence: $S \to w_1 \to \ldots \to w$, or $S \to^* w$. 
\end{definition}

\begin{definition}
The \textit{language specified by the context-free grammar} $G=\langle \Sigma, \mathcal{N}, P, S\rangle$ (denoted $\mathcal{L}(G)$) is a set of words derivable from the start nonterminal of the given grammar: $\mathcal{L}(G) = \{ w \mid S \to ^* w \}$.
\end{definition}

\begin{definition}\label{def:ebnf}
    A context-free grammar $G=\langle \Sigma, \mathcal{N}, P, S\rangle$ is in \textit{Extended Backus-Naur Form} (EBNF) if productions have the form $\mathcal{N} \to E$ where $E$ is a regular expression over $\mathcal{N} \cup \Sigma$.
\end{definition}

Grammar in EBNF can be converted Backus-Naur Form, but such transformation requires introducing new productions and new nonterminals that leads to a significant increase in the size of the grammar and, subsequently, poor performance of path querying algorithms.  

We use a generalization of a \textit{parse tree} to represent the result of a query.

\begin{definition}
    \textit{Parse tree} of the sequence $w$ w.r.t. the context-free grammar $G=\langle \Sigma, \mathcal{N}, P, S\rangle$ in EBNF is a rooted, node-labelled, ordered tree where: 
    \begin{itemize}
        \item All leaves are labelled with the elements of $\Sigma \cup \{\varepsilon\}$, where $\varepsilon \notin \Sigma \cup \mathcal{N}$ is a special symbol to demote the empty string. Left-to-right concatenation of leaves forms $w$.
        \item All other nodes are labelled with the elements of $\mathcal{N}$.
        \item The root is labelled with $S$.
        \item Let $w_0$ is an ordered concatenation of labels of node labelled with $N_i$. Then $w_0 \in \mathcal{L}(r)$, where $N_i \to r \in P$.  
    \end{itemize}
\end{definition}

A set of problems for the graph $D$ and the language $\mathcal{L}$ can be formulated: 
\begin{align}
R &= \{(u,v) \mid \text{exists a path } u\pi v \text{ in } D, \omega(\pi) \in \mathcal{L}  \} \label{eqn:all_pairs_reachability}
\\
Q &= \{\pi \mid \text{exists a path } u\pi v \text{ in } D, \omega(\pi) \in \mathcal{L}  \} \label{eqn:all_pairs_all_paths}
\\
R(I) &= \{(u,v) \mid \text{exists a path } u\pi v \text{ in } D, u \in I, \omega(\pi) \in \mathcal{L}  \} \label{eqn:multiple_source_reachability}
\\
Q(I) &= \{\pi \mid \text{exists a path } u\pi v \text{ in } D, u \in I, \omega(\pi) \in \mathcal{L}  \} \label{eqn:multiple_source_all_paths}
\end{align}

Here~\ref{eqn:all_pairs_reachability} is the \textit{all-pairs reachability problem}, and~\ref{eqn:all_pairs_all_paths} is the \textit{all-pairs all-paths problem}. The additional parameter $I$ denotes the set of start vertices in~\ref{eqn:multiple_source_reachability} and~\ref{eqn:multiple_source_all_paths}. Thus~\ref{eqn:multiple_source_reachability} and~\ref{eqn:multiple_source_all_paths} --- are the \textit{multiple source reachability} and the \textit{multiple source all-paths} problems respectively. Note, that for the \textit{all-paths} problems, the result $Q$ can be an infinite set. Typically, the algorithms for these problems build a finite structure which contains all paths of interest, not the explicit set of paths. As far as the intersection of a regular and a context-free language is context-free~\cite{MR0151376}, the mentioned finite representation of the result for reachability problem is a representation of a context-free language.

This work is focused on a context-free language. In some cases, the language can be regular, but not context-free. This corresponds to two specific cases of formal lqnguage path querying (FLPQ): \textit{Context-Free Path Querying} (CFPQ) and \textit{Regular Path Querying} (RPQ).  
\section{Recursive State Machines}\label{section:RSM}

\textit{Recursive state machine} or RSM~\cite{10.1145/1075382.1075387} is a way to represent context-free languages in a way resembling finite automata. It allows us to use a graph-based representation of context-free languages specification and unify the processing of regular and context-free languages. We will use the following definition of RSM in this work.

\begin{definition}
    \textit{Recursive state machine} (RSM) is a tuple $\mathcal{R} = \langle \mathcal{N},\Sigma,B,B_S,Q,Q_S\rangle$ where
    \begin{itemize}
       \item $N$ is a set of nonterminal symbols, 
       \item $\Sigma$ is a set of terminal symbols, 
       \item $Q$ is a set of states of the RSM, 
       \item $Q_S$ is a set of the start states of all \textit{boxes}, 
       \item $B=\{B_{N_i} \mid N_i \in \mathcal{N}\}$ is a set of boxes, where each \\ $B_{N_i} = \langle Q_{N_i}, q_S, Q_F^{N_i}, \delta \rangle$ is a deterministic finite automaton without $\varepsilon$-transitions called a \textit{box}. Here:  
        \begin{itemize}
            \item $Q_{N_i} \subseteq Q$ is a set of states of the box, 
            \item $q_S \in Q_{N_i} \cap Q_S$ is the start state of the box, 
            \item $Q_F^{N_i} \subseteq Q_{N_i}$ is a set of final states of the box, 
            \item $\delta \subseteq Q_{N_i} \times  (\Sigma\cup Q_S) \times Q_{N_i}$ is a transition function of the box.
        \end{itemize}
    \item $B_S \in B$ is the start box of the RSM.
    \end{itemize}
\end{definition}

Thus, RSM is a set of deterministic finite state machines without $\varepsilon$-transitions over the alphabet $\Sigma \cup Q_S$. But the computation process of RSM differs from the one of DFA because it involves using a stack to handle transitions labelled with elements of $Q_S$.

We can use RSM to compute the reachable vertices in a graph. To describe the current state of the computation, we use \textit{configurations}. Computation can be defined as a transition between configurations.

\begin{definition}
    A \textit{Configuration} $C_{\mathcal{R}}$ of the computation of the RSM $\mathcal{R}=\langle \mathcal{N},\Sigma,B,B_S,Q,Q_S \rangle$ over the graph $D=\langle V,E,L \rangle$ is a tuple $(q,v,\mathcal{S})$ where 
    \begin{itemize}
        \item $q \in Q$ is a current state of RSM,
        \item $\mathcal{S}$ is the current stack, whose frames have one of two types: 
        \begin{itemize} 
            \item return addresses frame (elements of $Q$) to specify states to continue computation after the call is finished;
            \item parsing tree node to store fragments of a parsing tree,
        \end{itemize}
        \item $v \in V$ is the current vertex (current position in the input).
    \end{itemize}
\end{definition}

\begin{definition}\label{def:rsm_transition}
    A \textit{transition step} of the RSM specifies how to get new configurations of RSM, given the current configuration. $C_{\mathcal{R}} \vdash W$ denotes that $\mathcal{R}$ can go to each configuration in $W$ from the configuration $C_{\mathcal{R}}$.
    \begin{align*}
    (q,v,w_0::s::\mathcal{S})  \vdash & \{ (q',v',t::w_0::s::\mathcal{S}) \mid (q,t,q') \in \delta, (v,t,v') \in E\} \\
                       & \cup \{(s', v, q'::w_0::s::\mathcal{S}) \mid (q,s',q') \in \delta, s' \in Q_S \} \\
                       & \cup \{(s,v,\textit{Node}(N_i, w_0)::\mathcal{S}) \mid q \in Q_F^{N_i}\}
    \end{align*}
    where $w_0$ is a possibly empty sequence of terminals and nonterminal nodes. 
\end{definition}

To simplify the acceptance condition, we introduce the concept of an \textit{extended RSM}. 

\begin{definition}
    For the given RSM $\mathcal{R}=\langle \mathcal{N},\Sigma,B,B_S,Q,Q_S\rangle$,
    the \textit{extended RSM} 
    $$\mathcal{R}'=\langle \mathcal{N} \cup \{S'\},\Sigma \cup\{\$\},B \cup {B'_S},B'_S,Q \cup \{q_0',q_1',q_2'\},Q_S \cup \{q_0'\}\rangle$$
    is an RSM which defines the same language and is built from $\mathcal{R}$ by adding a new start box 
    $$B'_S = \langle \{q_0',q_1',q_2'\}, q_0', \{q_2'\}, \{(q_0',q_0,q_1'),(q_1',\$, q_2')\} \rangle$$ where 
    \begin{itemize}
        \item $q_0$ is a start state of $B_S$,
        \item \$ is a special symbol to mark the end of input, $\$ \notin \Sigma$,
        \item $q_i'$ are newly added states, $q_i'\notin Q$.
    \end{itemize}
\end{definition} 

Finally, for the given extended RSM $\mathcal{R}$ and the given graph $D$ we say that $v_n$ is reachable from $v_0$  w.r.t. $\mathcal{R}$ if $(q'_0,v_0,[]) \vdash^* C$ such that $(q'_1,v_n,[N_S]) \in C$, where $\vdash^*$ denotes zero or more transition steps, and $N_S$ is a node for the start nonterminal of the original (not extended) RSM. Additionally, $N_S$ represents a respective path.

Now we need a way to compute transitions and to build trees efficiently, avoiding recomputation and infinite cycles that are possible with the naive implementation. Moreover, we need a compact representation of all paths of interest whose number can be infinite.

\subsection{Example}\label{section:example_of_rsm}

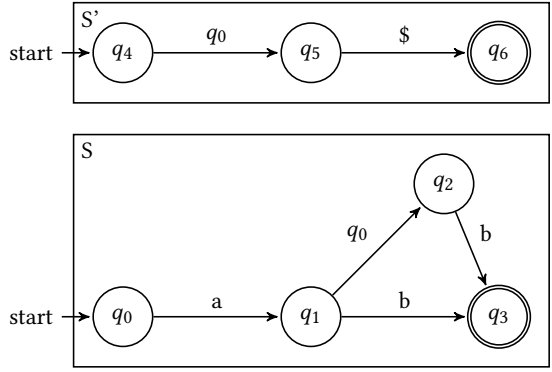
\begin{figure}
    \centering
    \begin{tikzpicture}[->,>=stealth',shorten >=1pt,auto,node distance=2.5cm, semithick]

  \node[initial,state]   (F)              {$q_4$};
  \node[state]           (G) [right of=F] {$q_5$};
  \node[accepting,state] (H) [right of=G] {$q_6$};
  \node[draw=black, fit= (F) (G) (H), inner sep=0.25cm] (J) {};
  \node[below right] at (J.north west) {S'};

  \path (F) edge              node {$q_0$} (G)
        (G) edge              node {$\$$} (H); 

  \node[initial,state] (A) [below = 2.7cm of F] {$q_0$};
  \node[state]         (B) [right of=A] {$q_1$};
  \node[state]         (D) [above right of=B] {$q_2$};
  \node[accepting,state]         (C) [right of=B] {$q_3$};
  \node [draw=black, fit= (A) (C) (D), inner sep=0.25cm] (E) {};
  \node[below right] at (E.north west) {S};

  \path (A) edge              node {a} (B)
        (B) edge              node {$q_0$} (D)
            edge              node {b} (C)
        (D) edge              node {b} (C);
\end{tikzpicture}
    
    \caption{Extended RSM for grammar $S \to a \ b \mid a \ S \ b$}
    \label{fig:example-rsm}
\end{figure}

\begin{figure}
    \centering

\begin{tikzpicture}[->,>=stealth',shorten >=1pt,auto,node distance=2.8cm, semithick]

  \node[state] (A)                    {$v_0$};
  \node[state]         (B) [right of=A] {$v_1$};
  
  \path (A) edge  [loop left] node {a} (A)
            edge  [bend left] node {b} (B)
        (B) edge  [bend left] node {b} (A);
\end{tikzpicture}

    \caption{Input graph}
    \label{fig:input-graph}
\end{figure}
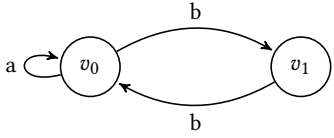

In this section, we introduce a step-by-step example of the context-free constrained path querying using RSM and the naive computation strategy. 

Suppose that the input is a graph $D$ presented in figure~\ref{fig:input-graph} and the grammar  $G$ has two productions $S \to a \ b \mid a \ S \ b$. The start vertex is $v_0$ and our goal to find at least one path to each reachable vertex (w.r.t. $G$). The extended RSM for the given grammar is presented in figure~\ref{fig:example-rsm}. 

The initial configuration is $(q_4,v_0,[])$: we start from the initial state of the box for $S'$, the initial position in the graph is a start vertex $v_0$, the stack is empty. In each step, we apply rules from definition~\ref{def:rsm_transition} to compute new configurations. The sequence of transitions, presented below, allows us to find a path $$v_0 \xrightarrow{a} v_0 \xrightarrow{b} v_1$$
from $v_0$ to $v_1$ (step~\ref{eq:naive-rsm-step-res-1}) and a path 
$$v_0 \xrightarrow{a} v_0 \xrightarrow{a} v_0 \xrightarrow{b} v_1 \xrightarrow{b} v_0$$
from $v_0$ to itself (step~\ref{eq:naive-rsm-step-res-2}). 

\begin{align}
\setcounter{equation}{0}
(q_4,v_0,[])     \vdash & \{(q_0,v_0,[q_5])\} \\
(q_0,v_0,[q_5])  \vdash & \{(q_1,v_0,[a,q_5])\} \\
(q_1,v_0,[a,q_5])\vdash & \{(q_0,v_0,[q_2,a,q_5]) \nonumber\\ 
                        & , (q_3,v_1,[b,a,q_5])\} \\
(q_0,v_0,[q_2,a,q_5]) \vdash & \{(q_1,v_0,[a,q_2,a,q_5])\} \\
(q_3,v_1,[b,a,q_5])   \vdash & \boxed{\{(q_5,v_1,[S(a,b)])\}} \label{eq:naive-rsm-step-res-1}\\
(q_1,v_0,[a,q_2,a,q_5]) \vdash & \{(q_3,v_1,[b,a,q_2,a,q_5]) \nonumber \\
                               & , (q_0,v_0,[q_2,a,q_2,a,q_5])\} \label{eq:naive-rsm-step-6} \\
(q_3,v_1,[b,a,q_2,a,q_5]) \vdash & \{(q_2,v_1,[S(a,b),a,q_5])\} \\
(q_2,v_1,[S(a,b),a,q_5]) \vdash & \{(q_3,v_0,[b,S(a,b),a,q_5])\} \\
(q_3,v_0,[b,S(a,b),a,q_5]) \vdash & \boxed{\{(q_5,v_0,[S(a,S(a,b),b)])\}} \label{eq:naive-rsm-step-res-2}
\end{align}

Note that we have no conditions to stop computation. In our example, we can continue computation and get new paths between $v_0$ and $v_1$. Moreover, there is an infinite number of such paths. Additionally, the selection of the next step is not deterministic. One can choose the configuration $(q_0,v_0,[q_2,a,q_2,a,q_5])$ to continue computations after step~\ref{eq:naive-rsm-step-6} with chance to fall into an infinite cycle, but choosing $(q_3,v_1,[b,a,q_2,a,q_5])$ we get a new path.
\section{Generalized LL for CFPQ}\label{section:GLL-CFPQ}

Our solution is based on the generalized LL parsing algorithm~\cite{SCOTT2010177} which was shown to generalize well to graph processing~\cite{Grigorev:2017:CPQ:3166094.3166104}. As a result of parsing, GLL can construct a \textit{Shared Packed Parse Forest} (SPPF)~\cite{SCOTT20131828} --- a special data structure which represents all possible derivations of the input in a compressed form. Alternatively, GLL can provide the result in the form of \textit{binary subtree sets}~\cite{SCOTT201963}. It was shown in~\cite{Grigorev:2017:CPQ:3166094.3166104} that SPPF can be naturally used to represent the result of the query for the \textit{all-paths} problem finitely (see \ref{eqn:all_pairs_all_paths} and~\ref{eqn:multiple_source_all_paths} in the set of problems).

In our work, we modify the GLL-based CFPQ algorithm to handle RSMs instead of grammars in BNF as a query specification. It enables performance improvement and native handling of both context-free and regular languages.

\subsection{Paths Representation}

Firstly, we need a way to represent a possibly infinite set of paths of interest. 
Note that the solution of all-paths problem is a set of paths which can be treated as a language using the $\omega$ function (\ref{eqn:omega}). This language is a context-free language because it is an intersection of a context-free and a regular language. Thus, the paths can be represented in a way a context-free language can be represented. 

The obvious way to represent a context-free language is to specify a respective context-free grammar. This way was used by Jelle Hellings in~\cite{10.1007/978-3-030-61133-0_7}: the query result is represented as a context-free grammar, and paths are extracted by generating strings.

Another way is a so-called \textit{path index}: a data structure which is constructed during query processing and allows one to reconstruct paths of interest with the additional traversal. This idea is used in a wide range of graph analysis algorithms, notably by Rustam Azimov et al in~\cite{10.1145/3461837.3464513}. In our case, the path index for the graph $D=\langle V,E,L \rangle$ and the query $\mathcal{R}=\langle \mathcal{N}, \Sigma, B, B_S, Q, Q_S \rangle$ is a $\mathcal{K} \times \mathcal{K}$ square matrix $\mathcal{I}$, where $\mathcal{K} = |Q|\cdot|V|$. Columns and rows are indexed by tuples of form $(q_i, v_j), q_i \in Q, v_j \in V$. Each cell $\mathcal{I}[(q_i,v_j),(q_l,v_k)]$ represents information about a \textit{range} $R^{q_i,v_j}_{q_l,v_k}$. \textit{Range} or a \textit{matched range} $R^{q_i,v_i}_{q_j,v_j}$ corresponds to the fact that there is a chain of transitions from the configuration $(q_i,v_i,\mathcal{S}_i)$ to the configuration $(q_j,v_j,\mathcal{S}_j)$, or $(q_i,v_i,\mathcal{S}_i) \vdash^* (q_j,v_j,\mathcal{S}_j)$. The symbol $\epsilon$ denotes the empty range: an empty sequence of configurations. Namely, $\mathcal{I}[(q_i,v_j),(q_l,v_k)]$ is a set which can contain the elements of three types:
\begin{itemize}
    \item $t \in \Sigma$ means that $(q_i,v_j,\mathcal{S}') \vdash (q_l,v_k,t::\mathcal{S}') $; 
    \item $N_m \in \mathcal{N}$ means that $(q_i,v_j,\mathcal{S}') \vdash^* (q_l,v_k,N_m::\mathcal{S}')$; 
    \item $I_{q_a,v_b}$ is an \textit{intermediate point} that is used to denote that the range $R^{q_i,v_j}_{q_l,v_k}$ was combined from two ranges $R^{q_i,v_j}_{q_a,v_b}$ and $R^{q_a,v_b}_{q_l,v_k}$. Equivalently, $$(q_i,v_j,\mathcal{S}') \vdash^* (q_a,v_b,\mathcal{S}'') \vdash^* (q_l,v_k,\mathcal{S}''').$$
\end{itemize}

A Shared Packed Parse Forest (SPPF) was proposed by Jan Rekers in~\cite{rekers1992parser} to represent parse forest without duplication of subtrees. Later, other versions of SPPF were introduced in different generalized parsing algorithms. It was also shown that SPPF is a natural way to compactly represent the structure of all paths which satisfy a CFPQ. In our work, we use SPPF with the following types of nodes. 

\begin{itemize}
    \item A \textit{Terminal node} to represent a matched edge label.
    \item A \textit{Nonterminal node} to represent the root of the subtree which corresponds to paths can be derived from the respective nonterminal.  
    \item An \textit{Intermediate node} which corresponds to the intermediate point used in the path index. This node has two children, both are range nodes.
    \item A \textit{Range node} which corresponds to a matched range. It represents all possible ways to get a specific range and can have arbitrary many children. A child node can be of any type, besides a Range node. Nodes of this type can be reused.
    \item An \textit{Epsilon node} to represent the empty subtree in the case when nonterminal produces the empty string. 
\end{itemize}

In this paper, we describe a version which creates the path index, and then we show how to reconstruct SPPF using this index. The version provided can be easily adopted to build SPPF directly during query evaluation, without creating the path index.

\subsection{GLL-Based CFPQ Algorithm}

In this section, we provide the detailed description of the GLL-based CFPQ algorithm that solves the \textit{multiple source all-paths} problem and handles queries in the form of RSM. We use the improved version of GLL, proposed by Afroozeh and Izmaylova in~\cite{10.1007/978-3-662-46663-6_5}, as a basis for our solution. 

Firstly, we introduce the basic components of the algorithm.

\begin{definition}
    A \textit{descriptor} is a 4-tuple $(q,v,s,r)$ where 
    \begin{itemize} 
      \item $q$ is a state of RSM, 
      \item $v$ is a position in the input graph, 
      \item $s$ is a pointer to the current top of the stack, 
      \item $r$ is a matched range.
    \end{itemize}
\end{definition}

A descriptor represents the state of the system completely; thus computation can be continued from the specified point without any additional information, only given a descriptor. There are two collections of descriptors in the algorithm. The first one is a collection $\mathcal{Q}$ of descriptors to handle. Each descriptor handles once, so the second one is a collection of already handled descriptors $\mathcal{Q}_{\text{handled}}$.

Graph-structured stack (GSS) represents a set of stacks in a compact graph-like form which enables reusing of the common parts of stacks.  We use an optimized version proposed by Afroozeh and Izmaylova in~\cite{10.1007/978-3-662-46663-6_5}. In this version, vertices identify calls, and edges contain return addresses and information about the matched part of the input. We slightly adapt GSS as follows:

\begin{definition}
    Suppose one has the graph $D=\langle V, E ,L \rangle$ and evaluates the query $\mathcal{R}=\langle \mathcal{N}, \Sigma, B, B_S, Q, Q_S \rangle$ over it. Then a \textit{Graph Structured Stack} (\textit{GSS}) is a directed edge-labelled graph $D_{\text{GSS}} = \langle V_{\text{GSS}}, E_{\text{GSS}}, L_{\text{GSS}} \rangle$, where 
    \begin{itemize}
        \item $V_{\text{GSS}} \subseteq Q \times V$,
        \item $E_{\text{GSS}} \subseteq V_{\text{GSS}} \times L_{\text{GSS}} \times V_{\text{GSS}}$, $L_{\text{GSS}} \subseteq Q \times R$,        
        \item $R \subseteq Q \times V \times Q \times V$ is a set of matched ranges.        
    \end{itemize}
\end{definition}

Two main operations over GSS are \textit{push} and \textit{pop}. Push adds a vertex and an edge if they do not already exist.
Doing pop, we go through all outgoing edges to get return addresses and to get the new stack head. There is no predefined order to handle descriptors. Thus, the new outgoing edge can be added at any time. To guarantee that all possible outgoing edges will be handled, we should save information about what pops have been done. In our case, we save the matched range of the current descriptor. When new outgoing edges are added to the GSS vertex, we should check whether this vertex has been popped and handle the newly added outgoing edges if necessary.

We suppose that the input of the algorithm is an extended RSM $\mathcal{R}' = \langle \mathcal{N}', \Sigma',B',B'_S,Q',Q'_S\rangle$ and graph $D=\langle V, E ,L \rangle$. Also note that we solve multiple-source CFPQ, so the set of start vertices $V_S \subseteq V$ is specified by the user.
The first step of the algorithm is the initialization. For each start vertex $v_i\in V_S$, we create a GSS vertex $s=(q'_0,v_i)$ and then create a descriptor $(q'_0,v_i,s,\epsilon)$, add it to $\mathcal{Q}$. Here $q'_0$ is the initial state of $B'_S$.

In the main loop, we handle descriptors one by one. Namely, while $\mathcal{Q}$ is not empty, we pick a descriptor $d = (q_0,v_0,s_0,r_0), r_0=R^{p_0,u_0}_{q_0,v_0}$, add it to $\mathcal{Q}_{\text{handled}}$, and process it. Processing consists of the following cases that provide a way to compute the transition step of RSM, as defined in Section~\ref{def:rsm_transition}, extended with path index creation.

Let 
$$
\Delta = \bigcup_{\substack{B_k \in B' \\ B_k = \langle Q^k, q^k_S, Q_F^k, \delta ^k \rangle}} \delta ^k.
$$
\begin{enumerate}
    \item \textbf{Handling of terminal transitions.} The terminal transitions are handled similarly to transitions in a finite automaton: for all matched terminals, we simultaneously move forward to the next state and the next vertex; the stack does not change; the matched range extends right with the matched symbol.
    Thus, the new set of descriptors is
    \begin{align}
    D_1 = \{&(q_1,v_1,s_0, R^{p_0,u_0}_{q_1,v_1}) \nonumber \\ 
            & \mid (v_0,t,v_1) \in E, (q_0,t,q_1) \in \Delta, t \in \Sigma \}.
    \end{align}
    This time, we add the terminal $t$ to $\mathcal{I}[(q_0,v_0),(q_1,v_1)]$ and the intermediate point $I_{q_0,v_0}$ to $\mathcal{I}[(p_0,u_0),(q_1,v_1)]$.
    
    \item \textbf{Handling of nonterminal transitions.}
    For each nonterminal symbol $N_0$ such that there exists a transition $(q_0,N_0,q_1)$, we should start processing of the nonterminal $N_0$. To achieve it, we should push the return address $q_1$ and the accumulated range $r_0$ to the stack. To do it, the vertex $s_1$ with label $(q_0,v_0)$ is created or reused. Next, we create the descriptor  $d = (q^{N_0}_{0},v_0,s_1,\epsilon)$, where $q^{N_0}_{0}$ is a start state of $B_{N_0}$. Additionally, we should handle the stored pops not to miss the newly added outgoing edge. For all stored matched ranges $R^{q_0,v_0}_{q_3,v_3}$ for the current GSS vertex $s_1$,  we create the range $R^{q_0,v_0}_{q1,v_3}$. Finally, we combine the new range $R^{p_0,u_0}_{q_1,v_3}$ from the range of the current descriptor, the newly created one, and the respective descriptor $(q_0, v_3,q',R^{p_0,u_0}_{q_1,v_3})$. 
    We also add intermediate points of created descriptors to $\mathcal{I}$.
    Finally,
    \begin{align}
    D_2 = \{& q^{N_0}_{0},v_0,s_1,\epsilon  \nonumber \\
            & \mid s_1 = (q_0,v_0), (q_0,N_0,q_1) \in \Delta, \nonumber \\  
            & \ \ \  q^{N_0}_{0}\text{ is a start state of }B_{N_0}\}  \nonumber \\
     \cup \{&q_0, v_3,q',R^{p_0,u_0}_{q_1,v_3} \nonumber \\ 
            & \mid s_1 = (q_0,v_0), (q_0,N_0,q_1), (q_0,N_0,q_1) \in \Delta,\nonumber \\ 
            & R^{q_0,v_0}_{q_3,v_3} \in s_1.\text{stored\_pops} \}.
    \end{align}
    \item \textbf{Handling of the final state}. If $q_0$ is a final state of $B_{N_0}$, it means that the recognition of the corresponding nontermininal is completed. So, we should pop from the stack to determine the point from which we restart the recognition and return the respective states to continue. Namely, for all outgoing edges from $s_0=(r_0,w_0)$ of form $(s_0,(q_1,R^{q_2,v_2}_{q_3,w_0}),s_1)$ we use $q_1$ as the return address and use the range from label to create the following ranges. First of all, we create the range $R^{q_3,w_0}_{q_1,v_0}$ that corresponds to the recognized nonterminal, and save $N_0$ to $\mathcal{I}[(q_3,w_0),(q_1,v_0)]$. Next, we combine this range with the range from GSS edge getting $R^{q_2,v_2}_{q_1,v_0}$, and store intermediate point $I_{q_3,w_0}$ to $\mathcal{I}[(q_2,v_2),(q_1,v_0)]$. Note, that we should store the information about these pops. 
    Finally
    \begin{align}
    D_3 = \{ &(q_1,v_0,s_1,R^{q_2,v_2}_{q_1,v_0}) \nonumber \\
             & \mid (s_0,(q_1,R^{q_2,v_2}_{q_3,w_0}),s_1) \in E_{\text{GSS}}, s_0 = (r_0,w_0)\}. 
    \end{align}
\end{enumerate}

Now, we have a set of the newly created descriptors $D_{\text{created}} = D_1 \cup D_2 \cup D_3$. We drop out the descriptors which have been already handled, add the rest to $\mathcal{Q}$, and repeat the main loop again.

The result of the algorithm is a path index $\mathcal{I}$, which allows one to check reachability or to reconstruct paths or SPPF by traversing starting from the cell corresponding to the range of interest.

\subsection{Example}

\begin{figure*}
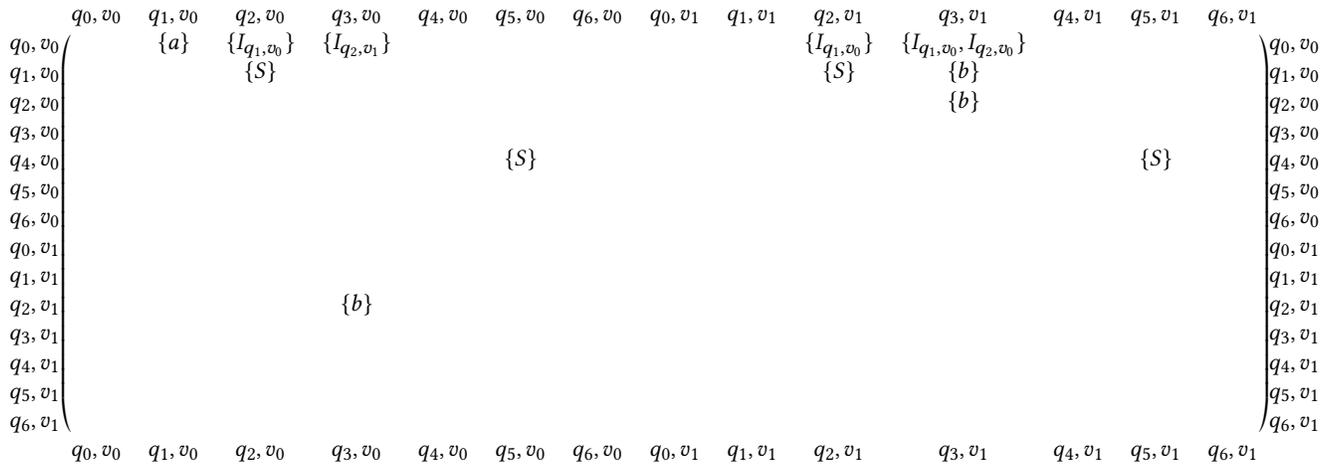

\setcounter{MaxMatrixCols}{30}
$\begin{pNiceMatrix}[first-row,last-row,first-col,last-col,nullify-dots]
        & q_0,v_0 & q_1,v_0 & q_2,v_0       & q_3,v_0       & q_4,v_0 & q_5,v_0 & q_6,v_0 & q_0,v_1 & q_1,v_1 & q_2,v_1         & q_3,v_1       & q_4,v_1 & q_5,v_1 & q_6,v_1          \\
q_0,v_0 &         & \{a\}   &\{I_{q_1,v_0}\}&\{I_{q_2,v_1}\}&         &         &         &         &         & \{I_{q_1,v_0}\} &\{I_{q_1,v_0}, I_{q_2,v_0}\}&         &         &         & q_0,v_0\\
q_1,v_0 &         &         &  \{S\}        &               &         &         &         &         &         &   \{S\}         & \{b\}         &         &         &         & q_1,v_0\\
q_2,v_0 &         &         &               &               &         &         &         &         &         &                 & \{b\}        &         &         &         & q_2,v_0\\
q_3,v_0 &         &         &         &               &         &         &         &         &         &                 &         &         &         &         & q_3,v_0\\
q_4,v_0 &         &         &         &               &         & \{S\}   &         &         &         &                 &         &         &\{S\}    &         & q_4,v_0\\
q_5,v_0 &         &         &         &               &         &         &         &         &         &                 &         &         &         &         & q_5,v_0\\
q_6,v_0 &         &         &         &               &         &         &         &         &         &                 &         &         &         &         & q_6,v_0\\
q_0,v_1 &         &         &         &               &         &         &         &         &         &                 &         &         &         &         & q_0,v_1\\
q_1,v_1 &         &         &         &               &         &         &         &         &         &                 &         &         &         &         & q_1,v_1\\
q_2,v_1 &         &         &         &   \{b\}       &         &         &         &         &         &                 &         &         &         &         & q_2,v_1\\
q_3,v_1 &         &         &         &         &         &         &         &         &         &         &         &         &         &         & q_3,v_1\\
q_4,v_1 &         &         &         &         &         &         &         &         &         &         &         &         &         &         & q_4,v_1\\
q_5,v_1 &         &         &         &         &         &         &         &         &         &         &         &         &         &         & q_5,v_1\\
q_6,v_1 &         &         &         &         &         &         &         &         &         &         &         &         &         &         & q_6,v_1\\
        & q_0,v_0 & q_1,v_0 & q_2,v_0 & q_3,v_0 & q_4,v_0 & q_5,v_0 & q_6,v_0 & q_0,v_1 & q_1,v_1 & q_2,v_1 & q_3,v_1 & q_4,v_1 & q_5,v_1 & q_6,v_1          \\
\end{pNiceMatrix}$
\caption{The path index created during the steps represented in table~\ref{tbl:gll-example}}
\label{fig:sppf-matrix}
\end{figure*}

Suppose that we have the same graph and RSM, as we have used in \ref{section:example_of_rsm} (\ref{fig:input-graph} and \ref{fig:example-rsm} respectively).
We construct a path index and then show how to create SPPF using it. 
All the steps of the index creation algorithm are presented in the table~\ref{tbl:gll-example}. The final index is presented in figure~\ref{fig:sppf-matrix}. Note that this example is rather simple, so we avoid handling the stored pops.

The first step is the initialization. The stack is initialized with one vertex which represents the initial call, while the range is empty.

In the second step, we start to handle the new nonterminal. So, the new vertex and the respective edge are added to GSS.

In the third step, we match the terminal $a$ and create the first record in the path index.

In the fourth step, we produce two descriptors. The first one has already been handled, so we should not process it again (thus, it is crossed out), but the creation of this descriptor leads to a new call. So, the new vertex and edge are added to GSS. Note that the respective vertex already exists, so only the new edge is added. Stack will not change any more after this step.

In the fifth step, two descriptors are created. One of them (boxed) denotes that we have recognized a path of interest. The second unique path is finished at the seventh step. All other paths are a combination of these two.

At the sixth step the terminal $b$ from the edge $v_1 \xrightarrow{b} v_0$ is handled. 

Finally, at the eight step, we get the empty set of descriptors to process $\mathcal{Q}$ and finish the process.

Next, we show how to reconstruct SPPF form path index. Suppose, we want to build SPPF for all paths that start in $v_0$, finish in $v_0$, and form words derivable from $S$. To achieve it, we start from the respective root range node $R^{q_4,v_0}_{q_5,v_0}$. The information about ways to build this range is stored in $\mathcal{I}[(q_4,v_0),(q_5,v_0)]$. Each element of the respective set becomes a child of the respective range node. In our case, we see that this range corresponds to the nonterminal node $S$. Thus, we should find all the ways to build range $R^{q_0,v_0}_{q_3,v_0}$ ($q_0$ is a start state for $S$ and $q_3$ is a final state). To do it, we look at $\mathcal{I}[(q_0,v_0),(q_3,v_0)]$. It contains the information about the intermediate point $I_{q_2,v_1}$ which means range $R^{q_0,v_0}_{q_3,v_0}$ has been built from the two ranges $R^{q_0,v_0}_{q_2,v_1}$ and $R^{q_2,v_1}_{q_3,v_0}$ that became children nodes  of $I_{q_2,v_1}$. Repeating this procedure allows us to build SPPF in a top-down direction. At some steps, for example when processing the intermediate point $I_{q_1,v_0}$, we found that some range nodes are always added to SPPF. Such nodes should be reused. The final result is presented in figure~\ref{fig:SPPF}.

Note that one can extract paths explicitly instead of SPPF construction in the similar fashion. 

\begin{table*}
\begin{tabular}{l c c c}
\hline \\
& 
Configuration
&
Path
&
Stack
\\
\hline \\
1 
&
$\vdash (q_4,v_0,(q_4,v_0),\epsilon)$ 
& 

&
\vspace{-0.2cm}
\begin{tikzpicture}[->,>=stealth',shorten >=1pt,auto,node distance=2.8cm, semithick]
  \node[state]         (A)              {$q_4,v_0$};
\end{tikzpicture} 
\\
\\
\hline \\
2 &
$(q_4,v_0,(q_4,v_0),\epsilon) \vdash \{(q_0,v_0,(q_0,v_0),\epsilon)\}$ 
&

&
\multirow{2}{*}[0.15cm]{
\begin{minipage}[t]{0.2\textwidth}
\begin{tikzpicture}[->,>=stealth',shorten >=1pt,auto,node distance=2.8cm, semithick]
  \node[state]         (A)              {$q_4,v_0$};
  \node[state]         (B) [right of=A] {$q_0,v_0$};
  \path (B) edge         node {$q_5, \epsilon $} (A);
\end{tikzpicture}
\end{minipage}
}
\\
3 & 
$(q_0,v_0,(q_0,v_0),\epsilon) \vdash \{(q_1,v_0,(q_0,v_0),R^{q_0,v_0}_{q_1,v_0})\}$ 
&
$[(q_0,v_0),(q_1,v_0)] \xleftarrow{} a$
&
\\
\\
\hline 
\\
4 
&
$(q_1,v_0,(q_0,v_0),R^{q_0,v_0}_{q_1,v_0}) \vdash \{\xcancel{(q_0,v_0,(q_0,v_0),\epsilon)}, (q_3,v_1,(q_0,v_0),R^{q_0,v_0}_{q_3,v_1})\}$ 
&
\begin{minipage}[t]{0.2\textwidth}
$\begin{array}{l}
\left[(q_1,v_0),(q_3,v_1)\right] \xleftarrow{} b \\ 
\left[(q_0,v_0),(q_3,v_1)\right] \xleftarrow{} (q_1,v_0) \\
\end{array}
$
\end{minipage}
&
\multirow{2}{*}[0.2cm]{
\begin{minipage}[t]{0.2\textwidth}
\begin{tikzpicture}[->,>=stealth',shorten >=1pt,auto,node distance=2.8cm, semithick]
  \node[state]         (A)              {$q_4,v_0$};
  \node[state]         (B) [right of=A] {$q_0,v_0$};
  \path (B) edge         node {$q_5, \epsilon$} (A)
        (B) edge  [loop] node[above]{$q_2, R^{q_0,v_0}_{q_1,v_0}$} (B) 
         ;
\end{tikzpicture}
\end{minipage}
}
\\

&
\\
5 
& 
$(q_3,v_1,(q_0,v_0),R^{q_0,v_0}_{q_3,v_1}) \vdash \{(q_2,v_1,(q_0,v_0),R^{q_0,v_0}_{q_2,v_1}), \boxed{(q_5,v_1,(q_4,v_0),R^{q_4,v_0}_{q_5,v_1})}\}$ 
&
\begin{minipage}[t]{0.2\textwidth}
$\begin{array}{l}
\left[(q_4,v_0),(q_5,v_1)\right] \xleftarrow{} S \\
\left[(q_1,v_0),(q_2,v_1)\right] \xleftarrow{} S \\
\left[(q_0,v_0),(q_2,v_1)\right] \xleftarrow{} (q_1,v_0) 
\end{array}
$
\end{minipage}

&
\\
\\
6 
& 
$(q_2,v_1,(q_0,v_0),R^{q_0,v_0}_{q_2,v_1}) \vdash \{(q_3,v_0,(q_0,v_0),R^{q_0,v_0}_{q_3,v_0})\}$ 
&
\begin{minipage}[t]{0.2\textwidth}
$\begin{array}{l}
\left[(q_2,v_1),(q_3,v_0)\right] \xleftarrow{} b \\
\left[(q_0,v_0),(q_3,v_0)\right] \xleftarrow{} (q_2,v_1)
\end{array}
$
\end{minipage}
&
\\
\\
7 
& 
$(q_3,v_0,(q_0,v_0),R^{q_0,v_0}_{q_3,v_0}) \vdash \{(q_2,v_0,(q_0,v_0),R^{q_0,v_0}_{q_2,v_0}), \boxed{(q_5,v_0,(q_4,v_0),R^{q_4,v_0}_{q_5,v_0})}\}$ 
&
\begin{minipage}[t]{0.2\textwidth}
$\begin{array}{l}
\left[(q_4,v_0),(q_5,v_0)\right] \xleftarrow{} S \\
\left[(q_1,v_0),(q_2,v_0)\right] \xleftarrow{} S \\
\left[(q_0,v_0),(q_2,v_0)\right] \xleftarrow{} (q_1,v_0)
\end{array}
$
\end{minipage}
&
\\
\\
8 
& 
$(q_2,v_0,(q_0,v_0),R^{q_0,v_0}_{q_2,v_0}) \vdash \{\xcancel{(q_3,v_1,(q_0,v_0),R^{q_0,v_0}_{q_3,v_1})}\}$ 
&
\begin{minipage}[t]{0.2\textwidth}
$\begin{array}{l}
\left[(q_0,v_0),(q_3,v_1)\right] \xleftarrow{} (q_2,v_0) \\
\left[(q_2,v_0),(q_3,v_1)\right] \xleftarrow{} b
\end{array}
$
\end{minipage}
&
\\
\\

\hline

\end {tabular}
\caption{The steps of the GLL-based CFPQ executed on the graph represented in figure~\ref{fig:input-graph} and the query represented in figure~\ref{fig:example-rsm}}
\label{tbl:gll-example}
\end{table*}

\begin{figure}
\centering
\begin{tikzpicture}[->,>=stealth',shorten >=1pt,auto,node distance=1.5cm, semithick,scale=0.8, every node/.style={scale=0.8}]

  \node[state]         (A)              {$R^{q_4,v_0}_{q_5,v_0}$};
  \node[state]         (B) [below of=A] {$S$};
  \node[state]         (C) [below of=B] {$R^{q_0,v_0}_{q_3,v_0}$};
  \node[state]         (D) [below of=C] {$I_{q_2,v_1}$};
  \node[state]         (E) [below of=D] {$R^{q_0,v_0}_{q_2,v_1}$};
  \node[state]         (F) [right of=E] {$R^{q_2,v_1}_{q_3,v_0}$};
  \node[state]         (G) [below of=E] {$I_{q_1,v_0}$};
  \node[state]         (H) [below of=F] {$b$};
  \node[state]         (J) [below of=G] {$R^{q_0,v_0}_{q_1,v_0}$};
  \node[state]         (I) [right of=J] {$R^{q_1,v_0}_{q_2,v_1}$};
  \node[state]         (K) [below left of=J] {$a$};
  \node[state]         (L) [below right of=I] {$I_{q_1,v_0}$};
  \node[state]         (M) [below right of=L] {$S$};
  \node[state]         (N) [below of=M] {$R^{q_0,v_0}_{q_3,v_1}$};
  \node[state]         (O) [below of=N] {$I_{q_1,v_0}$};
  \node[state]         (P) [right of=O] {$I_{q_2,v_0}$};
  \node[state]         (R) [below of=P] {$R^{q_0,v_0}_{q_2,v_0}$};
  \node[state]         (S) [right of=R] {$R^{q_2,v_0}_{q_3,v_1}$};
  \node[state]         (T) [below of=O] {$R^{q_1,v_0}_{q_3,v_1}$};
  \node[state]         (U) [below of=T] {$b$};
  \node[state]         (V) [below of=S] {$b$};
  \node[state]         (W) [below of=R] {$I_{q_1,v_0}$};
  \node[state]         (X) [below right of=W] {$R^{q_1,v_0}_{q_2,v_0}$};
  \node[state]         (Y) [below right of=X] {$S$};
  
  \path (A) edge          (B)
        (B) edge   (C)
        (C) edge   (D)
        (D) edge   (E)
        (D) edge   (F)
        (E) edge   (G)
        (F) edge   (H)
        (G) edge   (J)
        (G) edge   (I)
        (J) edge   (K)
        (I) edge   (L)
        (L) edge   (M)
        (M) edge   (N)
        (N) edge   (O)
        (N) edge   (P)
        (O) edge   (J)
        (O) edge   (T)
        (P) edge   (R)
        (R) edge   (W)
        (P) edge   (S)
        (T) edge   (U)
        (S) edge   (V)
        (W) edge[in=270, out=220]   (J)
        (W) edge   (X)
        (X) edge   (Y)
        (Y) edge[in=0, out=89]   (C)
        
         ;
\end{tikzpicture}
    \caption{The SPPF restored from the path index represented in~\ref{fig:sppf-matrix} for all paths from $v_0$ to $v_0$}
    \label{fig:SPPF}
\end{figure}

\section{Data Set Description}\label{section:dataset}

In this section, we provide a description of graphs and queries used for evaluation of implemented algorithms. Also, we describe common evaluation scenarios and evaluation environment.

We evaluated our solution on both classical regular and context-free path queries to estimate the ability to use the proposed algorithm as a universal algorithm for a wide range of queries. 

\subsection{Graphs}

For the evaluation, we use a number of graphs from CFPQ\_Data\footnote{CFPQ\_Data is a public set of graphs and grammars for CFPQ algorithms evaluation. GitHub repository: \url{https://github.com/FormalLanguageConstrainedPathQuerying/CFPQ_Data}. Accessed: 27.09.2023.} data set. We selected a number of graphs related to RDF analysis.
A detailed description of the graphs, namely the number of vertices and edges and the number of edges labeled by tags used in queries, is in Table~\ref{tab:graphs_for_evaluation_rdf}. Here ``bt'' is an abbreviation for  \textit{broaderTransitive} relationship.


\begin{table}
    \centering
    \scalebox{0.85}{
    \rowcolors{2}{black!2}{black!10}
    \begin{tabular}{| l | c | c | c | c | c | c |}
         \hline
         Graph name & $|V|$ & $|E|$ & \#subClassOf & \#type & \#bt\\
         \hline
         \hline
         Core               & 1 323     & 2 752      & 178        & 0        & 0 \\
         Pathways           & 6 238     & 12 363     & 3 117      & 3 118     & 0 \\
         Go\_hierarchy       & 45 007    & 490 109    & 490 109    & 0         & 0 \\
         Enzyme             & 48 815    & 86 543     & 8 163      & 14 989    & 8 156\\
         Eclass             & 239 111   & 360 248    & 90 962     & 72 517    & 0 \\
         Geospecies         & 450 609   & 2 201 532  & 0          & 89 065    & 20 867 \\
         Go                 & 582 929   & 1 437 437  & 94 514     & 226 481   & 0 \\
         Taxonomy           & 5 728 398 & 14 922 125 & 2 112 637  & 2 508 635 & 0 \\
         \hline
    \end{tabular}
    }
    \caption{RDF graphs for evaluation: number of vertices and edges, and number of edges with specific label}
    \label{tab:graphs_for_evaluation_rdf}
\end{table}
    

For regular path queries evaluation we used only RDF graphs because code analysis related graphs contain only two types of labels. Regular queries over such  graph are meaningless.

\subsection{Regular Queries}

Regular queries were generated using well-established set of templates for RPQ algorithms evaluation. Namely, we use templates presented in Table 2 in~\cite{Pacaci2020RegularPQ} and in Table 5 in~\cite{Wang2019}. We select four nontrivial templates (that contains compositions of Kleene star and union) that expressible in Cypher syntax to be able to compare native Neo4j querying algorithm with our solution. Used templates are presented in equations~\ref{eqn:reg_1},~\ref{eqn:reg_2},~\ref{eqn:reg_3}, and~\ref{eqn:reg_4}. Respective path patterns expressed in Cypher are presented in equations~\ref{eqn:cypher_reg_1},~\ref{eqn:cypher_reg_2},~\ref{eqn:cypher_reg_3}, and~\ref{eqn:cypher_reg_4}. Note that while Cypher  power is limited, our solution can handle arbitrary RPQs. We generate one query for each template and each graph. The most frequent relations from the given graph were used as symbols in the query template.

\begin{minipage}{0.24\textwidth}
\begin{align}
\textit{reg}_1 = & (a \mid b)^* \label{eqn:reg_1}\\
\textit{reg}_2 = & a^* \cdot b^* \label{eqn:reg_2}
\end{align}
\end{minipage}
\begin{minipage}{0.24\textwidth}
\begin{align}
\textit{reg}_3 = & (a \mid b \mid c)^+ \label{eqn:reg_3}\\
\textit{reg}_4 = & (a \mid b)^+ \cdot (c \mid d)^+ \label{eqn:reg_4}
\end{align}
\end{minipage}

\begin{align}
\textit{reg}_1^{\text{N4j}} = & \texttt{()-[:a | :b]->\{0,\}()} \label{eqn:cypher_reg_1}\\
\textit{reg}_2^{\text{N4j}} = & \texttt{()-[:a]->\{0,\}()-[:b]->\{0,\}()} \label{eqn:cypher_reg_2}\\
\textit{reg}_3^{\text{N4j}} = & \texttt{()-[:a | :b | :c]->\{1,\}()} \label{eqn:cypher_reg_3}\\
\textit{reg}_4^{\text{N4j}} = & \texttt{()-[:a | :b]->\{1,\}()-[:c | :d]->\{1,\}()} \label{eqn:cypher_reg_4}
\end{align}

Also note that exclude \textit{go\_hierarchy} graph from evaluation because it contains only one type of edges so it is impossible to express meaningful queries over it. 

\mycomment{
\begin{table}
    \centering
{\small
\renewcommand{\arraystretch}{1.2}
\rowcolors{2}{black!2}{black!10}
\begin{tabular}{|c|c||c|c|}
\hline

Name & Query & Name & Query \\
\hline
\hline
$Q_1$   & $a^*$                               & $Q_9^5$    & $(a \mid b \mid c \mid d \mid e)^+$                     \\
$Q_2$   & $a\cdot b^*$                        & $Q_{10}^2$ & $(a \mid b) \cdot c^*$                                  \\
$Q_3$   & $a \cdot b^* \cdot c^*$             & $Q_{10}^3$ & $(a \mid b \mid c)  \cdot d^*$                          \\
$Q_4^2$ & $(a \mid b)^*$                      & $Q_{10}^4$ & $(a \mid b \mid c \mid d)  \cdot e^*$                   \\
$Q_4^3$ & $(a \mid b \mid c)^*$               & $Q_{10}^5$ & $(a \mid b \mid c \mid d \mid e)  \cdot f^*$            \\
$Q_4^4$ & $(a \mid b \mid c \mid d)^*$        & $Q_{10}^2$ & $a \cdot b$                                             \\
$Q_4^5$ & $(a \mid b \mid c \mid d \mid e)^*$ & $Q_{11}^3$ & $a \cdot b \cdot c$                                     \\
$Q_5$   & $a \cdot b^* \cdot c$               & $Q_{11}^4$ & $a \cdot b \cdot c \cdot d$                             \\
$Q_6$   & $a^* \cdot b^*$                     & $Q_{11}^5$ & $a \cdot b \cdot c \cdot d \cdot f$                     \\
$Q_7$   & $a \cdot b \cdot c^*$               & $Q_{12}$   & $(a \cdot b)^+ \mid  (c \cdot d)^+$                     \\
$Q_8$   & $a? \cdot b^*$                      & $Q_{13}$   & $(a \cdot(b \cdot c)^*)^+ \mid  (d \cdot f)^+$          \\
$Q_9^2$ & $(a \mid b)^+$                      & $Q_{14}$   & $(a \cdot b \cdot (c \cdot d)^*)^+  \cdot (e \mid f)^*$ \\
$Q_9^3$ & $(a \mid b \mid c)^+$               & $Q_{15}$   & $(a \mid b)^+ \cdot (c \mid d)^+$                       \\
$Q_9^4$ & $(a \mid b \mid c \mid d)^+$        & $Q_{16}$   & $a \cdot b \cdot (c \mid d \mid e)$                     \\
\hline
\end{tabular}
}
\caption{Queries templates for RPQ evaluation}
\label{tbl:queries_templates}
\end{table}
}

\subsection{Context-Free Queries}

All queries used in our evaluation are variants of \textit{same-generation query}. For the \textit{RDF} graphs we use the same queries as in the other works for CFPQ algorithms evaluation: $G_1$~(\ref{eqn:g_1}), $G_2$~(\ref{eqn:g_2}), and $Geo$~(\ref{eqn:geo}). The queries are expressed as context-free grammars where $S$ is a nonterminal, \textit{subClassOf, type, broaderTransitive, }$ \overline{\textit{subClassOf}}$, $\overline{\textit{type}}$, $\overline{\textit{broaderTransitive}}$ are terminals or the labels of edges. We denote the inverse of an $x$ relation and the respective edge as $\overline{x}$.

\begin{align}
\label{eqn:g_2}
S \to \overline{\textit{subClassOf}} \ \ S \ \textit{subClasOf} \mid \textit{subClassOf}
\end{align}

\begin{align}
\begin{split}
\label{eqn:g_1}
S \to & \overline{\textit{subClassOf}} \ \ S \ \textit{subClasOf} \mid \overline{\textit{type}} \ \ S \ \textit{type}\\   & \mid \overline{\textit{subClassOf}} \ \ \textit{subClassOf} \mid \overline{\textit{type}} \ \textit{type}
\end{split}
\end{align}

\begin{align}
\begin{split}
\label{eqn:geo}
S \to & \textit{broaderTransitive} \ \  S \ \overline{\textit{broaderTransitive}} \\
      & \mid \textit{broaderTransitive} \ \  \overline{\textit{broaderTransitive}}
\end{split}
\end{align}

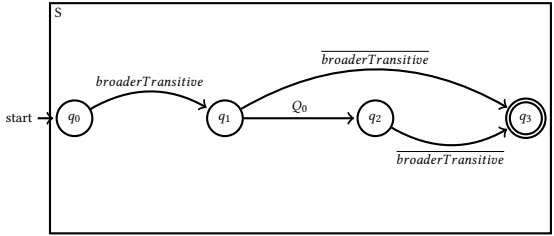
\begin{figure}
    \centering
    \begin{tikzpicture}[thick,scale=0.6, every node/.style={scale=0.6}, node distance=2cm,shorten >=1pt,on grid,auto]
           \node[state, initial] (q_0)   {$q_0$};
           \node[state] (q_1) [right=of q_0] {$q_1$};
           \node[state] (q_2) [right=of q_1] {$q_2$};
           \node[state, accepting] (q_3) [right=of q_2] {$q_3$};
           \node [draw=black, fit=(q_0) (q_1) (q_3), inner sep=2.3cm] (E) {};
           \node[below right] at (E.north west) {S};
           \path[->]
            (q_0) edge[bend left, above]  node {$broaderTransitive$} (q_1)
            (q_1) edge  node {$Q_0$} (q_2)
            (q_2) edge[bend right, below]  node {$\overline{broaderTransitive}$} (q_3)
            (q_1) edge[bend left, above]  node {$\overline{broaderTransitive}$} (q_3);
        \end{tikzpicture}
    \caption{RSM for query \textit{Geo} (\ref{eqn:geo})}
    \label{fig:rsm_geo}
\end{figure}

Respective RSM-s are presented in figures~\ref{fig:rsm_g1} for $G_1$,~\ref{fig:rsm_g2} for $G_2$, and~\ref{fig:rsm_geo} for \textit{Geo}.

\begin{figure}
    \centering
    \begin{tikzpicture} [thick,scale=0.6, every node/.style={scale=0.6}, node distance=2cm,shorten >=1pt,on grid,auto]
           \node[state, initial] (q_0)   {$q_0$};
           \node[state] (q_1) [above right=of q_0] {$q_1$};
           \node[state] (q_2) [right=of q_1] {$q_2$};
           \node[state, accepting] (q_3) [below right=of q_2] {$q_3$};
           \node[state] (q_4) [below right=of q_0] {$q_4$};
           \node[state] (q_5) [right=of q_4] {$q_5$};
           \node [draw=black, fit=(q_0) (q_1) (q_5) (q_3), inner sep=2.3cm] (E) {};
           \node[below right] at (E.north west) {S};
           \path[->]
            (q_0) edge[bend left, left]  node {$\overline{subClassOf}$} (q_1)
            (q_1) edge  node {$q_0$} (q_2)
            (q_2) edge[bend left, right]  node {$subClassOf$} (q_3)
            (q_1) edge[left]  node {$subClassOf$} (q_3)
            (q_0) edge[bend right, left] node {$\overline{type}$} (q_4)
            (q_4) edge  node {$q_0$} (q_5)
            (q_5) edge[bend right, right]  node {$type$} (q_3)
            (q_4) edge[left]  node {$type$} (q_3);
        \end{tikzpicture}
    \caption{RSM for query $G_1$ (\ref{eqn:g_1})}
    \label{fig:rsm_g1}
\end{figure}

\begin{figure}
    \centering
    \begin{tikzpicture}[thick,scale=0.6, every node/.style={scale=0.6}, node distance=2cm,shorten >=1pt,on grid,auto]
           \node[state, initial] (q_0)   {$q_0$};
           \node[state] (q_1) [right=of q_0] {$q_1$};
           \node[state] (q_2) [right=of q_1] {$q_2$};
           \node[state, accepting] (q_3) [right=of q_2] {$q_3$};
           \node [draw=black, fit=(q_0) (q_1) (q_3), inner sep=2.3cm] (E) {};
           \node[below right] at (E.north west) {S};
           \path[->]
            (q_0) edge[bend right, below]  node {$\overline{subClassOf}$} (q_1)
            (q_1) edge  node {$q_0$} (q_2)
            (q_2) edge[bend right, below]  node {$subClassOf$} (q_3)
            (q_0) edge[bend left, above]  node {$subClassOf$} (q_3);
        \end{tikzpicture}
    \caption{RSM for query $G_2$ (\ref{eqn:g_2})}
    \label{fig:rsm_g2}
\end{figure}

\subsection{Scenarios Description}

We evaluate the proposed solution on the \textbf{multiple sources reachability} scenario. We assume, that the size of the starting set is significantly less than the number of the input graph vertices. This limitation looks reasonable in practical cases. 

The starting sets for the multiple sources querying are generated from all vertices of a graph with a random permutation. We use chunks of size 1, 10, 100. For graphs with less than 10~000 vertices all vertices were used for querying, for graphs with from 10~000 to 100~000 vertices 10\% of vertices were considered starting ones, and for the graphs with more than 100~000 vertices only 1\% of vertices were considered. We use the same sets for all cases in all experiments to be able to compare results.

To check the correctness of our solution and to force the result stream, we compute the number of reachable pairs for each query.

\subsection{Evaluation Environment}

We run all experiments on the PC with Ubuntu 20.04 installed. It has Intel Core i7-6700 CPU, 3.4GHz, 4 threads (hyper-threading is turned off), and DDR4 64Gb RAM. We use OpenJDK 64-Bit Server VM Corretto-17.0.8.8.1 (build 17.0.8.1+8-LTS, mixed mode, sharing). JVM was configured to use 55Gb of heap memory: both \texttt{xms} and \texttt{xmx} are set to 55Gb. 

We use Neo4j 5.12.0. Almost all configurations of Neo4j are default. We only set  \texttt{memory\_transaction\_global\_max\_size} to \texttt{0}, which means unlimited memory usage per transaction.

As a competitor for our implementation, we use linear algebra based solution, integrated to RedisGraph by Arseniy Terekhov et al and described in~\cite{DBLP:conf/edbt/TerekhovPAZG21} and we use the configuration described in it for RedisGraph evaluation in our work.

\section{Performance of GLL on Queries in BNF and EBNF}\label{section:BFN_or_RSM}

As discussed above, different ways to specify context-free grammars can be used to specify the query. The basic one is BNF (see definition~\ref{def:bnf}), the more expressive (but not more powerful) is EBNF (see definition~\ref{def:ebnf}). EBNF not only more expressive, but potentially allows one to improve performance of query evaluation because avoid stack usage by replacing some recursive rules with Kleene star. RSM's allows one to natively represent grammars in EBNF and can be handled by GLL as described in section~\ref{section:GLL-CFPQ}.

We implemented and evaluated two versions of GLL-based CFPQ algorithm: one operates with grammar in BNF, another one operates with grammar in EBNF and utilizes RSM to represent it. At this step we use simple data structures graph representation and query representation, tuned for our algorithm. Both versions were evaluated in reachability only mode to estimate performance difference and to choose the best one to integrate with Neo4j. 

\mycomment{
\subsection{Implementation Details}
The modification of the GLL algorithm to operate with the grammar in EBNF is based on replacing \emph{positions in the grammar} in the descriptor with the state of a recursive automaton.
Accordingly, the transition in a context-free grammar is replaced with a transition in a recursive automaton.

Our modification, like the classic GLL algorithm, can be generalized to graphs. At the transition step in the recursive automaton, all transitions from the current state of the recursive automaton are processed along with all outgoing edges of the current vertex of the graph.

To make the implementation code itself as concise and understandable as possible for further extensions and optimizations, the Kotlin was chosen as a programming language for implementing.
}

\subsection{Evaluation}\label{sectioin:cfg_vs_rsm}

To assess the applicability of the proposed solution, we evaluate it on a number of real-world graphs and queries described in section~\ref{section:dataset}.

We compare performance of evaluation of queries in \textbf{reachability-only} mode with different sizes of start vertex set to estimate speedup of RSM-based version relative to BNF-based one.

The experimental study was conducted as follows.   
\begin{itemize}
    \item For all graphs, queries and start vertices sets, that described in section~\ref{section:dataset}, we measure evaluation time for both versions.
    \item Average time for each start vertices set size was calculated. So, for each graph, query, and start vertices set size we have an average time of respective query evaluation.
    \item Speedup as a ratio of BNF-based version evaluation time to RSM-based version evaluation time was calculated.
\end{itemize}

\begin{figure}
    \centering
    \includegraphics[width=0.47\textwidth]{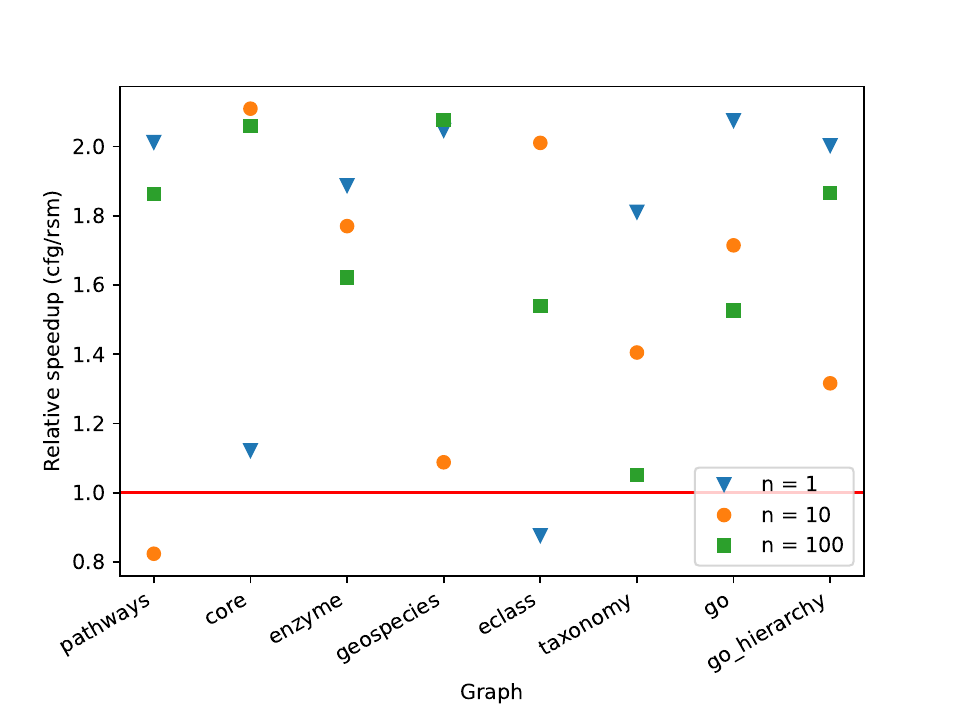}
    \caption{Multiple sources CFPQ reachability speedup (RSM over CFG) on RDF graphs and $G_1$}
    \label{fig:speedup-g1}
\end{figure}

\begin{figure}
    \centering
    \includegraphics[width=0.47\textwidth]{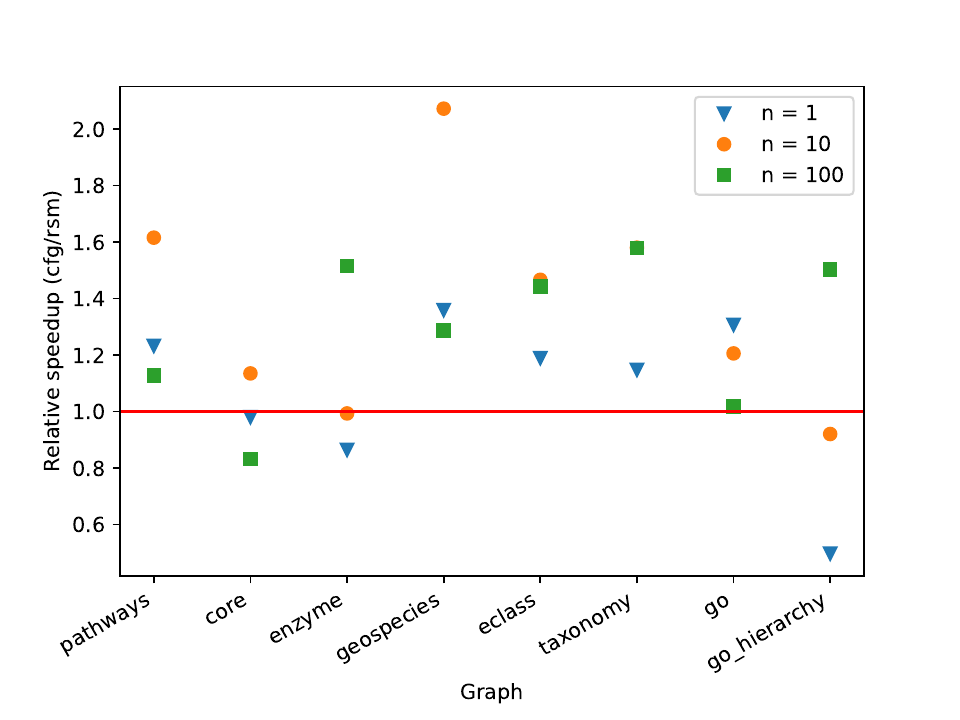}
    \caption{Multiple sources CFPQ reachability speedup (RSM over CFG) on RDF graphs and $G_2$}
    \label{fig:speedup-g2}
\end{figure}

\begin{figure}
    \centering
    \includegraphics[width=0.47\textwidth]{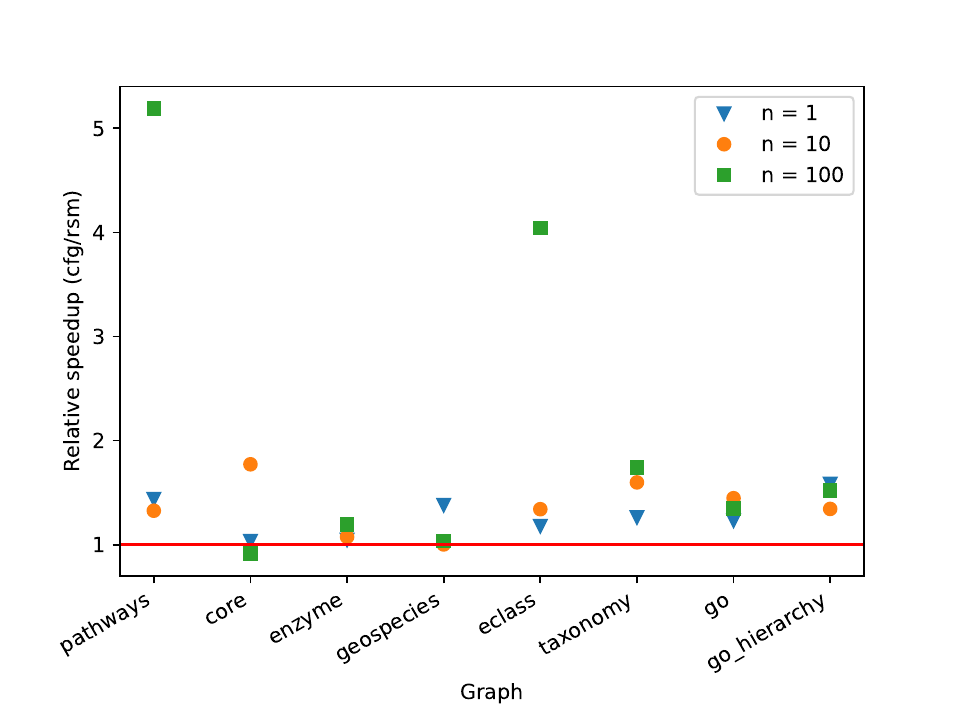}
    \caption{Multiple sources CFPQ reachability speedup (RSM over CFG) on RDF graphs and \textit{Geo}}
    \label{fig:speedup-geo}
\end{figure}

Results presented in figures~\ref{fig:speedup-g1},~\ref{fig:speedup-g2}, and~\ref{fig:speedup-geo} for queries $G_1$, $G_2$ and \textit{Geo} respectively. 
We can see that in almost the all cases RSM-based version is faster than BNF-based one (speedup is more than $1$). While in most cases speedup is not greater than 2, in some cases it can be more than 5 (see figure~\ref{fig:speedup-geo}: graph \textit{pathways}, grammar \textit{Geo}). Average speedup over all graphs and grammars  is 1.5. So we can conclude that RSM-based GLL demonstrates better performance than BNF-based one in average.

\section{CFPQ For Neo4J}\label{section:cfpq_fro_neo4j}

In this section we provide details on integration of GLL-based CFPQ to Neo4j graph database. We choose RSM-based version because our comparison (see section~\ref{sectioin:cfg_vs_rsm}) shows that it is faster than BNF-based one.

Also, we provide results of implemented solution evaluation which shows that, first of all, provided solution faster then similar linear algebra based solution for RedisGraph. Also, we show that on RPQs our solution compatible with the Neo4j native one and in some cases evaluates queries successfully while native solution fails with OutOfMemory exception.

\subsection{Implementation Details}

Neo4j stored procedure is a mechanism through which query language can be extended by writing custom code in Java in such way that it can be called directly via Cypher. 

We implemented Neo4j stored procedure which solves the reachability problem for the given set of start vertices and the given query. The procedure can be called as follows:
$$\texttt{CALL cfpq.gll.getReachabilities(nodes, q)}$$
where \texttt{nodes} is a collection of start nodes, and \texttt{q} is a string representation (or description) of RSM specified over relations types. Result of the given procedure is a stream of reachable pairs of nodes. Note that expressive power of our solution is limited: we can not use full power of Cypher inside our constraints. For example, we can not specify constrains on vertices inside our constraints.

\mycomment{
\begin{align*}
StartState(&id=0,nonterminal=Nonterminal(S),\\
           &isStart=true,isFinal=false) \\
State(&id=1,nonterminal=Nonterminal(S), \\
      &isStart=false,isFinal=false) \\
State(&id=4,nonterminal=Nonterminal(S),isStart=false,isFinal=false) \\
State(&id=3,nonterminal=Nonterminal(S),isStart=false,isFinal=true) \\
State(&id=2,nonterminal=Nonterminal(S),isStart=false,isFinal=false) \\
State(&id=5,nonterminal=Nonterminal(S), \\
      &isStart=false,isFinal=false) \\
TerminalEdge(&tail=0,head=1,terminal=Terminal(subClassOf\_r)) \\
TerminalEdge(&tail=0,head=4,terminal=Terminal(type\_r)) \\
TerminalEdge(&tail=1,head=3,terminal=Terminal(subClassOf)) \\
NonterminalEdge(&tail=1,head=2,nonterminal=Nonterminal(S)) \\
TerminalEdge(&tail=4,head=3,terminal=Terminal(type)) \\
NonterminalEdge(&tail=4,head=5,nonterminal=Nonterminal(S)) \\
TerminalEdge(&tail=2,head=3,terminal=Terminal(subClassOf)) \\
TerminalEdge(&tail=5,head=3,terminal=Terminal(type))
\end{align*}
}

We implemented a wrapper for Neo4j. Communication with the database is done using the Neo4j Native Java API. So, we used embedded database, which means it is run inside of the application and is not used as an external service.

Along with the existing modifications of GLL we made a Neo4j-specific one. Neo4j return result should be represented as a \texttt{Stream} and it is important to prevent early stream forcing, thus we changed all GLL internals to ensure that. This also has an added benefit that the query result is a stream, and thus it is possible to get the results on demand. 

\subsection{Evaluation}

To assess the applicability of the proposed solution we evaluate it on a number of real-world graphs and queries. To estimate relative performance we compare our solution with matrix-based CFPQ algorithm implemented in RedisGraph by Arseniy Terekhov et al in~\cite{DBLP:conf/edbt/TerekhovPAZG21}. Also, we compare performance of evaluation of queries in \textbf{reachability-only} mode on regular path queries with native Neo4j solution.

\begin{figure}
    \centering
    \includegraphics[width=0.47\textwidth]{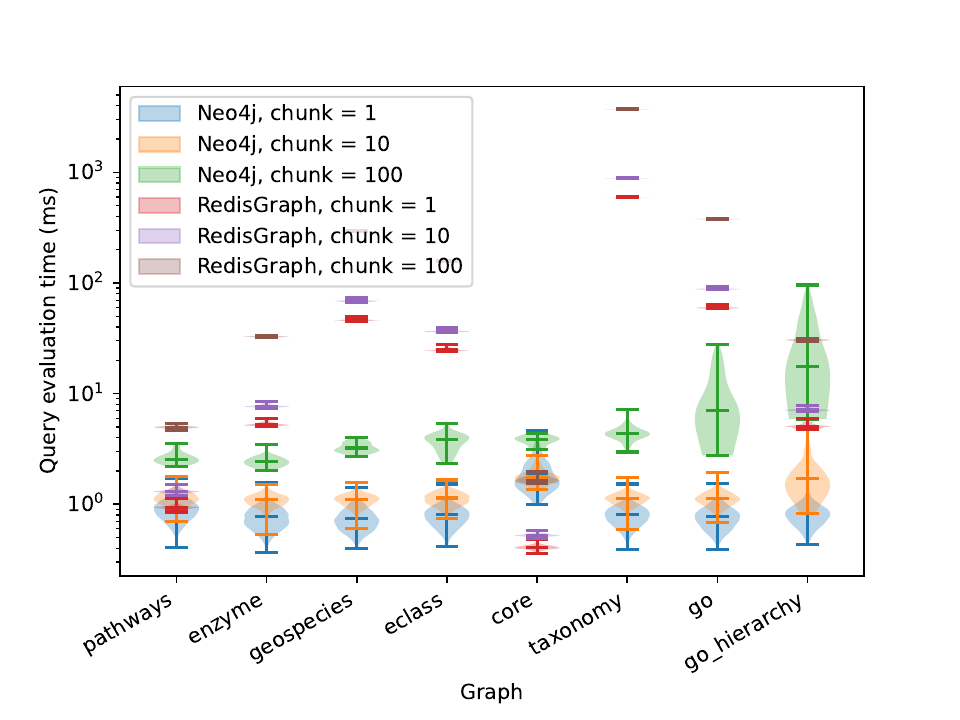}
    \caption{Multiple sources CFPQ reachability results for queries related to RDF analysis and $G_1$}
    \label{fig:performance-g1}
\end{figure}
\begin{figure}
    \centering
    \includegraphics[width=0.47\textwidth]{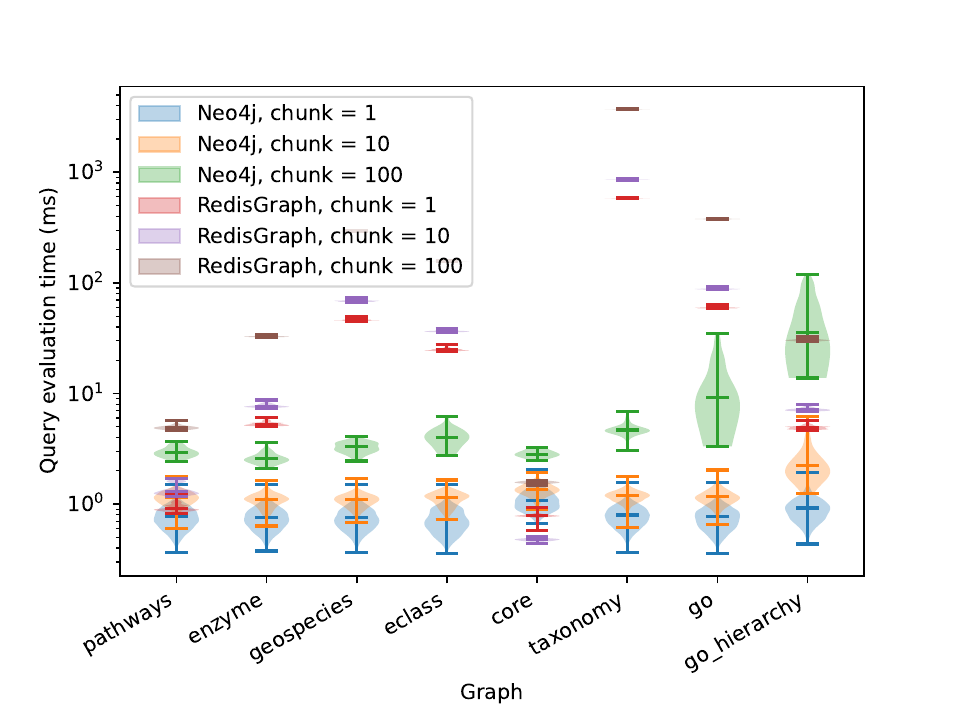}
    \caption{Multiple source CFPQ reachability results for queries related to RDF analysis and $G_2$}
    \label{fig:performance-g2}
\end{figure}
\begin{figure}
    \centering
    \includegraphics[width=0.47\textwidth]{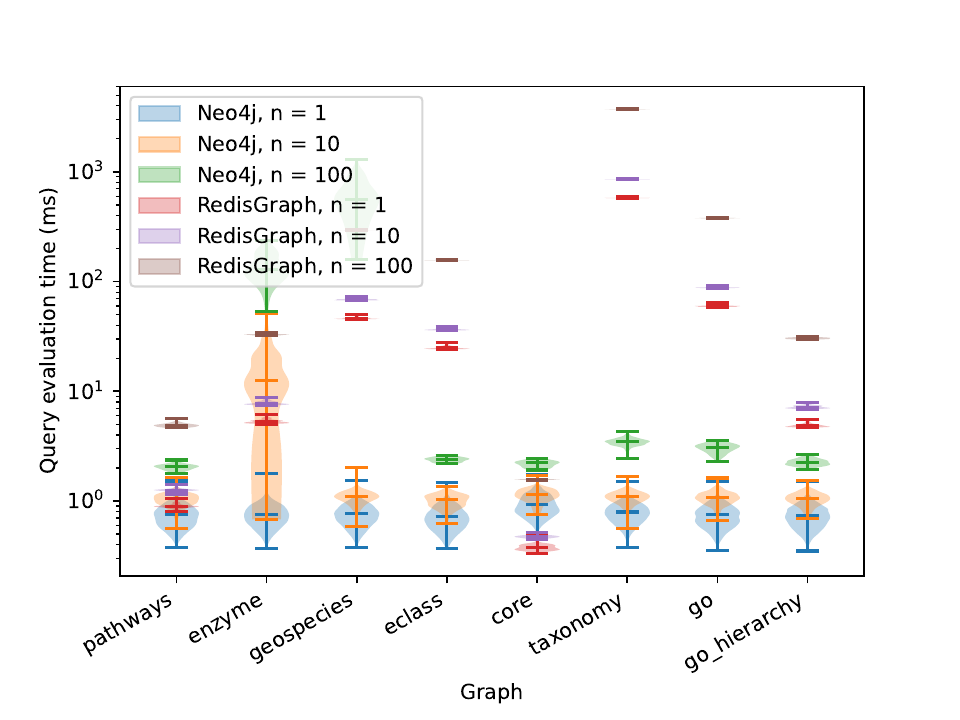}
    \caption{Multiple source CFPQ reachability results for queries related to RDF analysis and \textit{Geo}}
    \label{fig:performance-geo}
\end{figure}

The results of context-free path queries evaluation are presented in figures~\ref{fig:performance-g1} for $G_1$,~\ref{fig:performance-g2} for $G_2$, and~\ref{fig:performance-geo} for \textit{Geo}. 

The results show that query evaluation time depends not only on a graph size or its sparsity, but also on an inner structure of the graph. For example, the relatively small graph \textit{go\_hierarchy} fully consists of edges used in queries $G_1$ and $G_2$, thus evaluation time for these queries is significantly bigger than for some bigger but more sparse graphs, for example, for \textit{eclass} graph. Especially for relatively big start vertex set. Note, that the creation of relevant metrics for CFPQ queries evaluation time prediction is a challenging problem by itself and should be tackled in the future.

Also, we can see that in almost the all cases the proposed solution significantly faster than RedisGraph (in orders of magnitude in some cases). At the same time, in some cases (see results for graph \textit{core} and all queries) RedisGraph outperforms our solution. Also, one can see, that evaluation time for RedisGraph is more predictable. For our solution in some cases execution time highly depends on start set. For example, see figure~\ref{fig:performance-geo}, graph \textit{enzyme}.    

The particular important scenario is the case when the start set is a single vertex. 
The results of the \texttt{single-source reachability} show that such queries are reasonably fast: median time is about few milliseconds for all graphs and all queries. Note that even for single source queries, evaluation time highly depends on graph structure: evaluation time on \textit{core} graph significantly bigger then for all other graphs for all queries. Note, that \textit{core} is a smallest graph in terms of number of nodes and edges. Again, to provide an reliable metric to predict query execution time is a nontrivial task. Moreover, time grows with the size of a chunk, as expected. 

\begin{figure}
    \centering
    \includegraphics[width=0.47\textwidth]{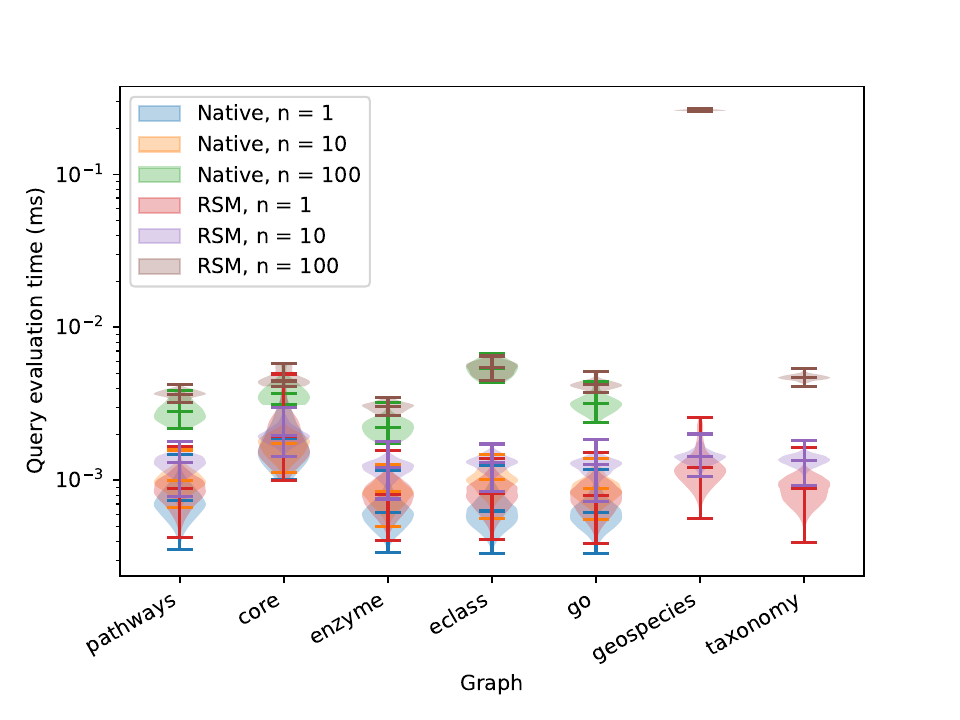}
    \caption{Multiple sources RPQ reachability results for queries related to RDF analysis and $\textit{reg}_1$ (native solution failed with OOM on last two graphs)}
    \label{fig:performance-reg1}
\end{figure}

\begin{figure}
    \centering
    \includegraphics[width=0.47\textwidth]{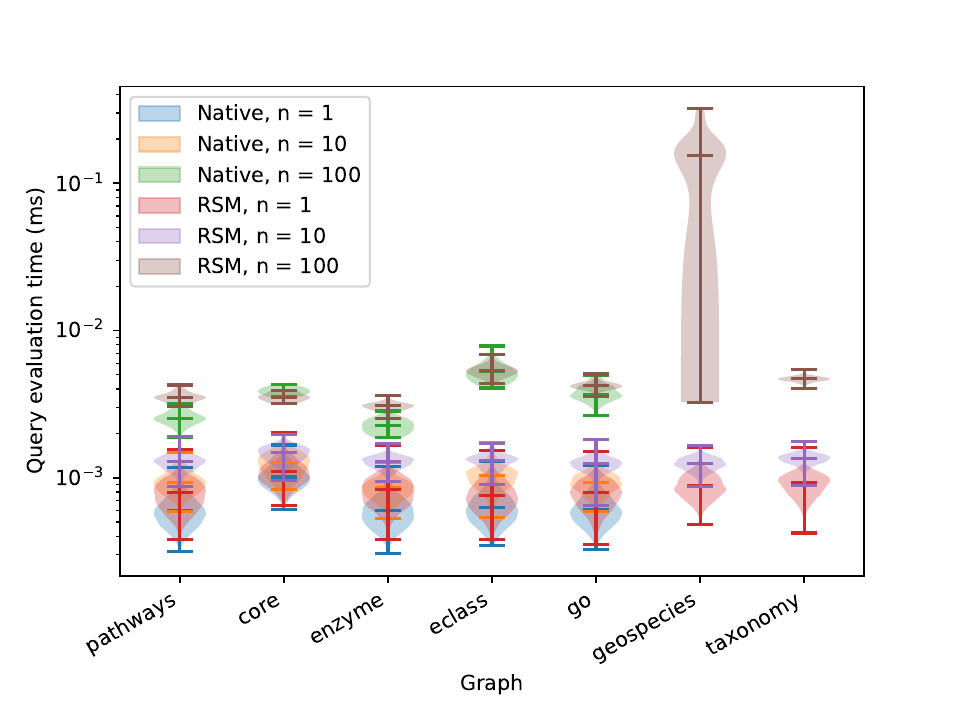}
    \caption{Multiple source RPQ reachability results for queries related to RDF analysis and $\textit{reg}_2$ (native solution failed with OOM on last two graphs)}
    \label{fig:performance-reg2}
\end{figure}

Partial results for RPQ evaluation are presented in figures~\ref{fig:performance-reg1} and~\ref{fig:performance-reg2} for $\textit{reg}_1$ (defined in ~\ref{eqn:reg_1}) and $\textit{reg}_2$ (defined in ~\ref{eqn:reg_2}) respectively. For queries $\textit{reg}_3$ (defined in ~\ref{eqn:reg_3}) and $\textit{reg}_4$ (defined in ~\ref{eqn:reg_4}) we get similar results. Note, that on \textit{geospecies} and \textit{taxonomy} graphs native solution failed with OutOfMemory exception, while our solution evaluates queries successfully. 

We can see that the proposed solution slightly slower than native Neo4j algorithm, but not dramatically: typically less than two times. Moreover, in some cases our solution comparable with native one (see figure~\ref{fig:performance-reg1} and figure~\ref{fig:performance-reg2}, graph \textit{eclass}), and in some cases our solution faster than native one (see figure~\ref{fig:performance-reg2}, graph \textit{core}).

Finally, we conclude that not only the linear algebra based CFPQ algorithms can be used in real-world graph analysis systems: the proposed GLL-based solution outperforms linear algebra based one. Moreover, we show that the proposed solution can be used as a universal algorithm for both RPQ and CFPQ.
\section{Related Work}\label{section:related_work}

The idea to use context-free grammars as constraints in path finding problem in graph databases was introduced and explored by Mihalis Yannakakis in~\cite{10.1145/298514.298576}. A bit later, the same idea was developed by Thomas Reps et al. as a general framework for static code analysis~\cite{10.1145/199448.199462}. Further, this framework, called Context-Free Language Reachability (CFL-r), became one of the most popular and actively developed. The landscape analysis of the area was recently provided by Andreas Pavlogiannis in~\cite{10.1145/3583660.3583664}. In the context of graph databases, the most recent systematic comparison of different CFPQ algorithms was done by  Jochem Kuijpers et al. A set of CFPQ algorithms was implemented and evaluated using Neo4J as a graph storage. Results were presented in ~\cite{Kuijpers:2019:ESC:3335783.3335791}. It was concluded that the existing algorithms are not performant enough to be used for real-world data analysis.

Regarding graph databases, CFPQ was applied in such graph analysis related tasks as biological data analysis~\cite{SubgraphQueriesbyContextfreeGrammars}, data provenance~\cite{8731467}, hierarchy analysis in RDF data~\cite{MEDEIROS2022101089, 10.1007/978-3-319-46523-4_38}.

Multiple CFPQ algorithms are based on different parsing algorithms and techniques. For example, an approach based on parsing combinators was proposed by Ekaterina Verbitskaia et al. in~\cite{10.1145/3241653.3241655}. Several algorithms based on LL-like and LR-like techniques were developed by Ciro M. Medeiros et al. in~\cite{10.1145/3427081.3427087, MEDEIROS201975, MEDEIROS2022101089}. Also, CFPQ algorithms were investigated by Phillip Bradford in~\cite{Bradford2009, 8249039} and Charles B. Ward et al. in~\cite{4625871}. An algorithm based on matrix equations was proposed by Yuliya Susanina in~\cite{10.1145/3318464.3384400}.
Paths extraction problem was studied by Jelle Hellings in~\cite{10.1007/978-3-030-61133-0_7}.

A set of linear algebra-based CFPQ algorithms was developed, including all-paths and single-path versions, proposed by Rustam Azimov et al. in~\cite{10.1145/3461837.3464513} and~\cite{10.1145/3398682.3399163} respectively, Kronecker product-based algorithm~\cite{10.1007/978-3-030-54832-2_6} was proposed by Egor Orachev et al., and multiple-source CFPQ algorithm for RedisGraph was proposed by Arseniy Terekhov et al. in ~\cite{DBLP:conf/edbt/TerekhovPAZG21}. 

Recursive state machines were studied in the context of CFPQ in several papers, including~\cite{10.1007/978-3-030-54832-2_6} where Egor Orachev et al. use RSM to specify context-free constraints, Yuxiang Lei et al.~\cite{10.1145/3591233} propose to use RSM to specify path constraints, and~\cite{10.1145/1328438.1328460} where Swarat Chaudhuri proposes a slightly subcubic algorithm for reachability problem for recursive state machines --- the equivalent to CFPQ problem.

GLL was introduced by Elizabeth Scott and Adrian Johnstone in~\cite{SCOTT2010177}. A number of modifications of the GLL algorithm were further proposed, including the version which supports EBNF without its transformation~\cite{SCOTT2018120} and the version which uses binary subtree sets~\cite{SCOTT201963} instead of SPPF. The latest version simplifies the algorithm and avoids the overhead of explicit graph construction. Within it, the optimized and simplified OOP-friendly version of GLL was proposed by Ali Afroozeh and Anastasia Izmaylova in~\cite{10.1007/978-3-662-46663-6_5}.
\section{Conclusion and Future Work}\label{section:conclusion}

In this work, we presented the GLL-based context-free path querying algorithm for Neo4j graph database. The implementation is available on GitHub: \url{https://github.com/vadyushkins/cfpq-gll-neo4j-procedure}.

Our solution uses Neo4j for graph storage, but the query language should be extended to support context-free constraints to make it useful. Both the extension of Cypher and the integration of our algorithm with the query engine are non-trivial challenges left for future work.

While GLL-based CFPQ potentially can be used to solve \textbf{all-paths} problem, currently we implemented the procedure for reachability problem only. Choosing useful strategies to enumerate paths and implementation of them is a direction for future research.  

The most important direction for future work is to find a way to provide an incremental version of the GLL-based CFPQ algorithm to avoid full query reevaluation when the graph is only slightly changed. While our solution is based on the well-known parsing algorithm and there are solutions for incremental parsing, development of the efficient incremental version of the GLL-based CFPQ algorithm is a challenging problem left for future research. 

Another direction is to create a parallel version of the GLL-based CFPQ algorithm to improve its performance on huge graphs. Although it seems natural to handle descriptors in parallel, the algorithm operates over global structures and the naive implementation of this idea leads to a significant overhead on synchronization.

\begin{acks}

We would like to thank Huawei Technologies Co., Ltd for supporting this research.

We thank Adrian Johnstone for pointing out the Generalized LL algorithm in our discussion at Parsing@SLE--2013, which motivated the development of the presented solution.

We thank George Fletcher for the discussion of evaluation of different CFPQ algorithms for Neo4j.


\end{acks}

\bibliographystyle{ACM-Reference-Format}
\bibliography{gll4graph}



\end{document}